\pdfoutput=1
\documentclass[longauth]{aa}

\usepackage{graphicx}
\usepackage{txfonts}

\usepackage{hyperref}

\usepackage{multirow}

\usepackage[binary-units]{siunitx}

\newcommand{\emm}[1]{\ensuremath{#1}}

\newcommand{\ch}[1]{\ensuremath{\mathrm{#1}}}

\newcommand{\HH}{\ch{H_2}}

\newcommand{\R}{\emm{\mathbb{R}}}

\newcommand{\paramNN}{\boldsymbol{\theta}} 

\newcommand{\boldx}{\mathbf{x}} 
\newcommand{\boldy}{\mathbf{y}} 
\newcommand{\boldm}{\mathbf{m}} 
\newcommand{\boldb}{\mathbf{b}} 
\newcommand{\boldW}{\mathbf{W}} 

\newcommand{\ffull}{\mathbf{f}} 
\newcommand{\fell}{f_\ell} 

\newcommand{\Pth}{\ensuremath{P_\text{th}}}
\newcommand{\Gnaught}{\ensuremath{G_\text{UV}}}
\newcommand{\AV}{\ensuremath{A_V^{\text{tot}}}}
\newcommand{\paramAngle}{\ensuremath{\alpha}}

\begin{document}

\title{Neural network-based emulation of interstellar medium models}

\author{%
  Pierre Palud\inst{\ref{LERMA/MEUDON},\ref{CRISTAL}}\thanks{Equal contribution.} %
  \and Lucas Einig\inst{\ref{IRAM},\ref{GIPSA-Lab}}\protect\footnotemark[1] %
  \and Franck Le Petit\inst{\ref{LERMA/MEUDON}} %
  \and Emeric Bron\inst{\ref{LERMA/MEUDON}} %
  \and Pierre Chainais\inst{\ref{CRISTAL}} %
  \and Jocelyn Chanussot\inst{\ref{GIPSA-Lab}} %
  \and J\'er\^ome Pety\inst{\ref{IRAM},\ref{LERMA/PARIS}} %
  \and Pierre-Antoine Thouvenin\inst{\ref{CRISTAL}} %
  \and David Languignon\inst{\ref{LERMA/MEUDON}} %
  \and Ivana Be\v{s}li\'c\inst{\ref{LERMA/PARIS}}
  \and Miriam G. Santa-Maria\inst{\ref{CSIC}} %
  \and Jan H. Orkisz\inst{\ref{IRAM}} %
  \and L\'eontine E. Ségal\inst{\ref{IRAM}, \ref{Toulon}} %
  \and Antoine Zakardjian\inst{\ref{IRAP}}
  \and S\'ebastien Bardeau\inst{\ref{IRAM}} %
  \and Maryvonne Gerin\inst{\ref{LERMA/PARIS}} %
  \and Javier R. Goicoechea\inst{\ref{CSIC}} %
  \and Pierre Gratier \inst{\ref{LAB}} %
  \and Viviana V. Guzman\inst{\ref{Catholica}} %
  \and Annie Hughes\inst{\ref{IRAP}} %
  \and François Levrier\inst{\ref{LPENS}} %
  \and Harvey S. Liszt\inst{\ref{NRAO}} %
  \and Jacques Le Bourlot\inst{\ref{LERMA/MEUDON}} %
  \and Antoine Roueff\inst{\ref{Toulon}} %
  \and Albrecht Sievers\inst{\ref{IRAM}} %
}


\institute{%
  LERMA, Observatoire de Paris, PSL Research University, CNRS, Sorbonne Universit\'es, 92190 Meudon, France. \email{pierre.palud@obspm.fr} \label{LERMA/MEUDON} %
  \and IRAM, 300 rue de la Piscine, 38406 Saint Martin d'H\`eres,  France, \email{einig@iram.fr} \label{IRAM} %
  \and Univ. Lille, CNRS, Centrale Lille, UMR 9189 - CRIStAL, 59651 Villeneuve d'Ascq, France, \label{CRISTAL} %
  \and Univ. Grenoble Alpes, Inria, CNRS, Grenoble INP, GIPSA-Lab, Grenoble, 38000, France. \label{GIPSA-Lab} %
  \and LERMA, Observatoire de Paris, PSL Research University, CNRS, Sorbonne Universit\'es, 75014 Paris, France. \label{LERMA/PARIS} %
  \and Instituto de Física Fundamental (CSIC). Calle Serrano 121, 28006, Madrid, Spain. \label{CSIC} %
  \and Université de Toulon, Aix Marseille Univ, CNRS, IM2NP, Toulon,
  France. \label{Toulon} %
  \and Institut de Recherche en Astrophysique et Planétologie (IRAP), Université Paul Sabatier, Toulouse cedex 4, France. \label{IRAP} %
  \and Laboratoire d'Astrophysique de Bordeaux, Univ. Bordeaux, CNRS,  B18N, Allee Geoffroy Saint-Hilaire,33615 Pessac, France. \label{LAB} %
  \and Instituto de Astrofísica, Pontificia Universidad Católica de Chile, Av. Vicuña Mackenna 4860, 7820436 Macul, Santiago, Chile. \label{Catholica} %
  \and Laboratoire de Physique de l'Ecole normale supérieure, ENS, Université PSL, CNRS, Sorbonne Université, Université de Paris, Sorbonne Paris Cité, Paris, France. \label{LPENS} %
  \and National Radio Astronomy Observatory, 520 Edgemont Road, Charlottesville, VA, 22903, USA. \label{NRAO} %
} %


\abstract
{
  The interpretation of observations of atomic and molecular tracers in the galactic and extragalactic interstellar medium (ISM) requires comparisons with state-of-the-art astrophysical models to infer some physical conditions.
  Usually, ISM models are too time-consuming for such inference procedures, as they call for numerous model evaluations.
  As a result, they are often replaced by an interpolation of a grid of precomputed models.
}
{
  We propose a new general method to derive faster, lighter, and more accurate approximations of the model from a grid of precomputed models for use in inference procedures.
}
{
  These emulators are defined with artificial neural networks (ANNs) with adapted architectures and are fitted using regression strategies instead of interpolation methods.
  The specificities inherent in ISM models need to be addressed to design and train adequate ANNs.
  Indeed, such models often predict numerous observables (e.g., line intensities) from just a few input physical parameters and can yield outliers due to numerical instabilities or physical bistabilities and multistabilities.
  We propose applying five strategies to address these characteristics:
  1)  an outlier removal procedure;
  2) a clustering method that yields homogeneous subsets of lines that are simpler to predict with different ANNs;
  3) a dimension reduction technique that enables us to adequately size the network architecture;  %
  4) the physical inputs are augmented with a polynomial transform to ease the learning of nonlinearities; and
  5) a dense architecture to ease the learning of simpler relations between line intensities and physical parameters.
}
{
  We compare the proposed ANNs with four standard classes of interpolation methods, nearest-neighbor, linear, spline, and radial basis function (RBF), to emulate a representative ISM numerical model known as the Meudon PDR code.
  Combinations of the proposed strategies produce networks that outperform all interpolation methods in terms of accuracy by a factor of 2 in terms of the average error (reaching 4.5\% on the Meudon PDR code) and a factor of 3 for the worst-case errors (33\%).
  These networks are also $1\,000$ times faster than accurate interpolation methods and require ten to forty times less memory.
}
{
  This work will enable efficient inferences on wide-field multiline observations of the ISM.
}


\maketitle


\section{Introduction}
\label{sec:introduction}

Many aspects of star and planet formation are still only partially understood.
Studies around the efficiency of star formation  require a better understanding of the effects of feedback mechanisms and of gas dynamics, both in the Milky Way and other galaxies.
In addition, understanding the evolution of interstellar matter from diffuse clouds to planet-forming disks requires investigations of the interstellar chemistry, for instance, examining the development of the chemical complexity or the fractionation of isotopologues.
New and large hyperspectral surveys in radioastronomy stand as a game-changer for the study of these processes, as they enable observing full molecular clouds ($\sim 10$ pc size) at a dense-core scale ($< 0.1$ pc) spatial resolution.
For instance, the “Orion B” IRAM-30m Large Program \citep{petyAnatomyOrionGiant2017} covers about $250$ pc$^2$ of the Orion B giant molecular cloud.
It has produced a hyperspectral
image of one million pixels and $200\,000$ spectral channels, allowing for the emission of dozens of molecules to be mapped over the whole cloud.
More generally, instruments with multispectral or hyperspectral capabilities such as the IRAM-30m, ALMA, NOEMA, and the James Webb spatial telescope (JWST) are now poised to provide observation maps with hundreds or thousands of pixels in multiple emission lines.

Astrophysical codes for interstellar medium (ISM) environments are able to model observed regions and link numerous observables (e.g., line intensities) to a few local physical conditions (e.g., the gas density or thermal pressure).
For instance, radiative transfer and excitation codes can be used to relate gas density, temperature, and column densities of detected species to their observable line intensities.
Such codes include RADEX~\citep{van_der_tak_computer_2007},
RADMC-3D~\citep{dullemond_radmc-3d_2012},
LIME~\citep{brinch_lime_2010},
MCFOST~\citep{pinte_mcfost_2022},
and MOLPOP-CEP~\citep{asensio_ramos_molpop-cep_2018}.
Some other codes adopt a more holistic approach and take multiple physical phenomena into account as well as their coupling, for instance, large chemical networks, thermal balance, and radiative transfer. Furthermore,
H\textsc{ii} region models such as Cloudy~\citep{ferland_2017_2017} reconstruct the chemical structure of ionized regions.
They evaluate line intensities from input parameters including illuminating star properties, the medium density, metallicity, and elementary abundances.
Shock models such as the Paris-Durham code~\citep{godard_models_2019} and the MAPPINGS code~\citep{sutherland_mappings_2018} compute the chemical structure of interstellar shocks and observables such as line intensities.
Here, the main input parameters are the shock velocity, pre-shock densities, and the intensity of the magnetic field.
Finally, photodissociation region (PDR) models such as the Meudon PDR code~\citep{le_petit_model_2006} describe the ultraviolet (UV) irradiated medium at the edge of molecular clouds in star-forming regions or diffuse interstellar clouds.
They compute the thermal and chemical structure of these objects as well as observables such as the atomic and molecular line intensities.
The input parameters mainly include the intensity of the incident stellar UV radiation field, the gas density or thermal pressure, the visual extinction, the metallicity and the cosmic ray ionization rate.
In the following, we use the term ``physical parameters'' to refer to a subset of interest of the input parameters that a code uses to compute observables.

For each of these models, small changes in the physical parameters can lead to very different predicted observables.
The adjustment of the physical parameters to allow the predicted observables to match the actual observations can therefore be used to estimate these physical parameters.
Codes that model the observed environment more realistically lead to more meaningful estimations.
However, the complexity of the physics considered in a code directly impacts its evaluation time and, hence, its applicability.

On the one hand, a simple 0D code such as RADEX can run in just a few seconds.
Such fast codes can be used directly for inference in minimization-based or Bayesian Markov chain Monte Carlo (MCMC) sampling approaches~\citep[chapter 7]{robert_monte_2004}, which require numerous iterative evaluations.
For instance, RADEX and UCLCHEM~\citep{holdshipUCLCHEMGasgrainChemical2017} have already been used as is in inference with Bayesian methods in low-dimensional cases~\citep{makrymallisUnderstandingFormationEvolution2014,holdshipBayesianInferenceRates2018,keilUCLCHEMCMCMCMCInference2022,behrensTracingInterstellarHeating2022,gratierNewReferenceChemical2016,maffucciAstrochemicalKineticGrid2018}.

On the other hand, a more comprehensive model such as the Meudon PDR code, which handles multiple physical processes on a 1D spatial grid, typically requires several hours of computations.
These durations are prohibitively long for inferring the physical parameters on large observation maps.
Such cases can be addressed by deriving a faster emulator either of the numerical model or of the full likelihood function, which includes both the numerical model and a noise model for observations.
For instance, the Bayesian algorithm BAMBI~\citep{graffBAMBIBlindAccelerated2012}, used for instance in~\cite{johannesson_bayesian_2016}, relies on the \textsc{SkyNet} neural network~\citep{graffSKYNETEfficientRobust2014} to emulate the full likelihood function.
Emulating the full likelihood requires the assumption of a noise model and it is therefore either observation-specific or generic. For instance, \textsc{SkyNet} assumes a Gaussian likelihood with a fixed variance for continuous variable inference.
In this work, we focus on full numerical code emulation to be able to apply the obtained emulator to any observation from any telescope, with any noise model and any set of lines.

In practice, the emulation of a numerical model is based on a grid of precomputed models that spans the relevant parameter space, generated prior to any comparison with observations.
A search for the point in the grid that best reproduces the observations is sometimes performed~\citep{sheffer_pdr_2011,sheffer_pdr_2013,joblin_structure_2018}.
A better and more common way of exploiting the grid is to approximate the numerical model using interpolation methods, which permits the observables for new points to be predicted with a lower evaluation time~\citep{wu_constraining_2018,ramambason_inferring_2022}.
In the following, a numerical code emulator defined from a grid of precomputed models using, for example, an interpolation method, is called a ``surrogate model.''
A grid of precomputed models is called a ``dataset.''

Interpolation methods have become the main approach to build surrogates of comprehensive ISM models over the last years thanks to their conceptual and implementation simplicity~(e.g., \citealt{wu_constraining_2018,ramambason_inferring_2022}).
Nearest-neighbor interpolation, linear interpolation, spline interpolation, and radial basis function (RBF) interpolation are the four most commonly used families of methods.
By definition, a surrogate model defined with an interpolation method passes exactly through the points of the dataset.
This constraint does not guarantee a good level of accuracy with respect to new points.
Besides, a surrogate model defined with an interpolation method requires the whole dataset to be stored, which can be very heavy if it contains many precomputed models or many quantities associated to each model.
Finally, although they are generally faster than the original numerical codes, interpolation methods handle outputs (i.e., observables) independently.
Thus, they are quite slow when the number of outputs is large.

In this work, we aim to derive accurate, fast, and light surrogate models.
To do so, we relaxed the constraint of having the model pass through the points of the dataset.
In this case, deriving a surrogate model thus becomes a regression problem, which benefits from many recent advances in numerical optimization developed for machine learning.
Such approaches have already been applied in ISM studies.
For instance, in~\cite{smirnov-pinchukovMachineLearningacceleratedChemistry2022}, a $k$-nearest-neighbor regression algorithm was used to emulate a protoplanetary disks model, while in~\cite{bronTracersIonizationFraction2021}, a random forest was trained to emulate a chemistry model.
However, most often, the versatile class of artificial neural networks (ANNs) is preferred to address the complexity of comprehensive ISM models.
For instance, ANN emulators of astrochemical models were derived in~\citep{demijollaIncorporatingAstrochemistryMolecular2019,holdshipChemulatorFastAccurate2021,grassi_reducing_2022}.
In addition, in~\cite{grassi_robo_2011}, the authors derived a new simulation code and an associated ANN emulator.
Here, we emulate a state-of-the-art ISM code, namely, the Meudon PDR code~\citep{le_petit_model_2006}.
Such a code sometimes produces outliers due to potential numerical instabilities or physical bistabilities or multistabilities.
It also predicts several thousands of observables from a handful of parameters, which is unusual in the machine learning community -- except for networks that generate structured data such as images, text, or times series.
In observations, only a fraction of these observables are measured.
However, different observations can detect very different subsets of observables.
To avoid having to derive one surrogate model per subset of observables, we chose to emulate the full code at once.

We present five main strategies to derive high performance surrogate models in these conditions.
First, a robust regression framework~\citep{rousseeuw_robust_1987} was used to identify and remove outliers.
Secondly, we applied a clustering method to derive homogeneous subsets of lines that are simpler to emulate, so that we could define and train one network per cluster.
We chose an adequate layer size in the network architecture thanks to a dimension reduction technique.
A polynomial transform of the input is applied to ease the learning of nonlinearities.
Finally, we limited any redundant computations with the recent dense architecture~\citep{huang_densely_2017}.
All obtained ANNs were then compared with interpolation methods with respect to speed, memory requirements, and accuracy.
The best obtained surrogate model will be exploited to perform inference of physical parameters from observations in~\cite{palud_estimating_nodate}.
We note that ANNs come with the ability to automatically and efficiently compute derivatives such as the gradient and the Hessian matrix, which enables using faster and more accurate inference methods.
All proposed ANNs were implemented using the PyTorch Python library\footnote{The code used to build the proposed ANNs can be found at \url{https://github.com/einigl/ism-model-nn-approximation}}.
The most accurate ANN obtained in this work has been made publicly available\footnote{\url{https://ism.obspm.fr/files/ArticleData/2023_Palud_Einig/2023_Palud_Einig_data.zip}}.

The paper is structured as follows.
Section~\ref{sec:nn} describes the emulation methods to be compared.
Section~\ref{sec:meudonpdr} describes the Meudon PDR code and the dataset of precomputed models.
It also introduces the framework used to compare surrogate models.
In Section~\ref{sec:using_nn}, we describe our design of ANNs that address the ISM numerical codes specificities.
In Section~\ref{sec:results}, we compare these ANNs with classic interpolation methods with respect to speed, memory requirements, and accuracy.
Section~\ref{sec:conclusion} provides our concluding remarks.


\section{Interpolation and regression methods}
\label{sec:nn}

Some notations used throughout this paper are introduced.
Four families of interpolation methods are then presented and feedforward ANNs succinctly described.
The regression paradigm is finally described with some ANNs specificities.
For a more detailed introduction on ANNs, we refer to~\citet[chapter 20]{shalev-shwartz_understanding_2014}\footnote{Accessible at \url{https://www.cs.huji.ac.il/~shais/UnderstandingMachineLearning/index.html}.}.

\subsection{Notation}

Throughout this paper, scalars are denoted with regular letters, such as indices $j$ or vector dimensions $D$ and $L$.
Vectors are denoted using bold lowercase letters, such as vectors of $D$ physical parameters $\boldx \in \R^D$, or vectors of $L$ line intensities $\boldy \in \R^L$.
Vectors are considered as collections of scalars: \mbox{$\boldy = (y_\ell)_{\ell=1}^L$} with, for example, $y_\ell$ as the intensity of a line $\ell$.
Matrices are written with bold uppercase letters.
The notation for functions is set accordingly, such as \mbox{$\ffull(\boldx) = (\fell(\boldx))_{\ell=1}^L$, with $\fell(\boldx)$} as the function that links an input $\boldx$ to the intensity of a line $\ell$.

\subsection{Interpolation methods in ISM studies}
\label{subsec:interpolation}

\begin{figure}[t]
  \centering
  \includegraphics[width=0.9\linewidth]{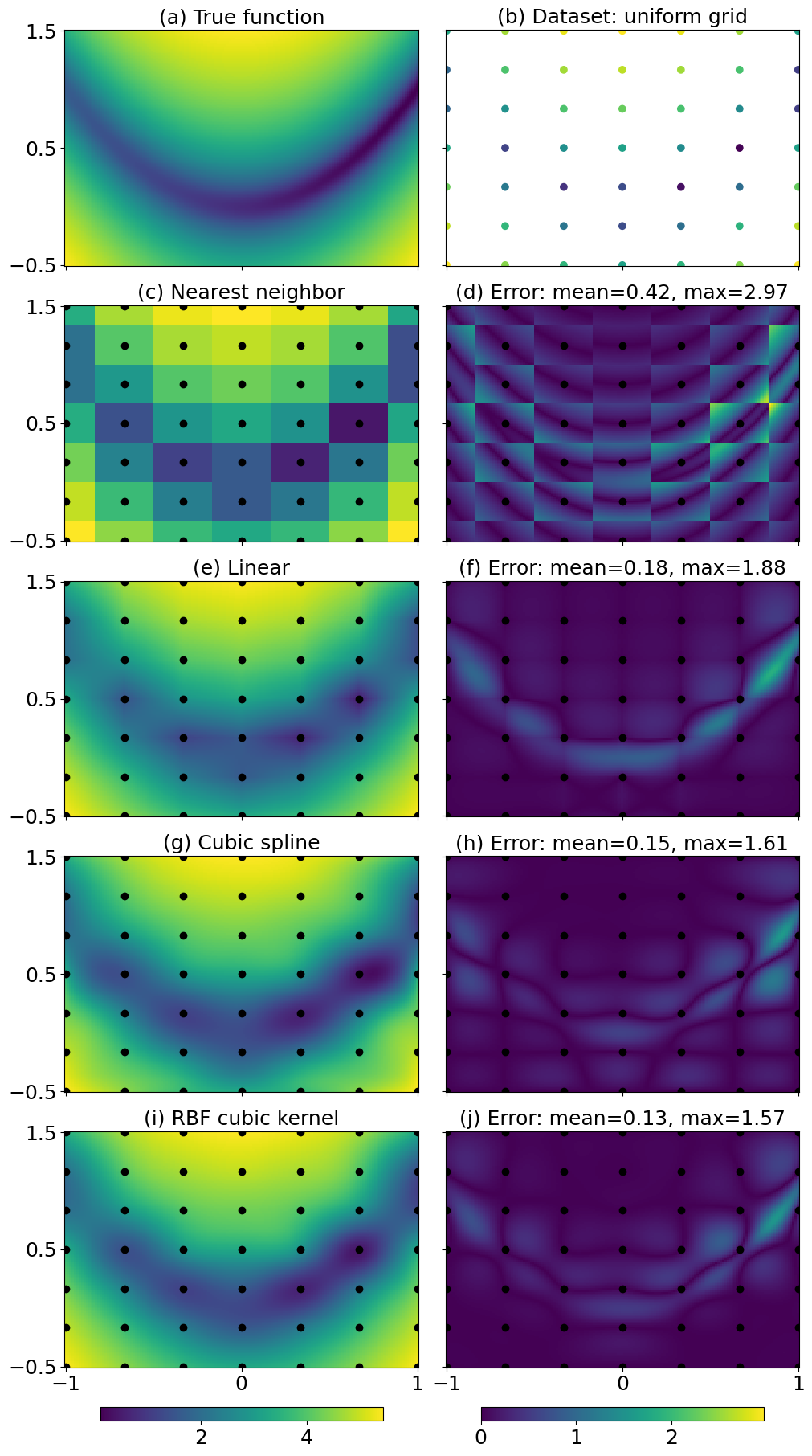}
  \caption{
    Comparison of the most popular interpolation methods on the log Rosenbrock function with a dataset structured as a coarse regular grid.
    (a)~Log Rosenbrock function $\log R$~(Eq.~\ref{eq:rosenbrock}), i.e., the true function that interpolation methods are to emulate.
    (b)~Coarse uniform grid and corresponding values of the true function.
    All four interpolation methods are fitted using these values only.
    The grid is also shown on the remaining figures.
    (c), (e), (g), and (i)~Surrogate model obtained with each interpolation algorithm.
    (d), (f), (h), and (j)~Absolute error~(Eq.~\ref{eq:def_ae}) between the corresponding surrogate model and the true function.
  }
  \label{fig:interpolation}
\end{figure}

Interpolation methods yield functions that pass exactly through the points of a dataset of precomputed models.
In this paper, four common families of interpolation methods are studied: nearest-neighbor interpolation, linear interpolation, spline interpolation~\citep{bojanov_spline_1993}, and RBF interpolation~\citep[chapter 6]{fasshauer_meshfree_2007}.
The nearest-neighbor interpolation assigns  the value of the closest point in the dataset to a new point.
It is fast but generally performs poorly in terms of accuracy.
It is somewhat equivalent to a search for the closest point in a grid, which is common in ISM studies~(e.g., \citealt{sheffer_pdr_2011,sheffer_pdr_2013,joblin_structure_2018}).
The piece-wise linear interpolation generally performs better, while remaining quite fast.
It first triangulates the dataset, so that a new point is associated to a cell of the triangulation.
It then returns a weighted average of the cell point values.
It was used in some ISM studies, such as in~\cite{ramambason_inferring_2022}.
Spline interpolation methods are based on piece-wise polynomials, yielding an even more accurate and still fast surrogate model.
Finally, the RBF interpolation, used, for example, in~\citet{wu_constraining_2018} to study a PDR, exploits the full dataset for each evaluation.
For a new point, it returns a weighted sum of the values of all the dataset points, where the weights depend on the distance to this new point.
Surrogate models defined with RBF interpolation are generally very accurate but slower than other interpolation methods.

In ISM studies, datasets of precomputed models are often structured as uniform grids \citep[i.e., as lattices,][]{joblin_structure_2018, wu_constraining_2018}.
This structure is not necessary for RBF interpolation methods or in regression approaches, and other structures that can be obtained with, for example, Latin hypercube sampling~\citep{mckayComparisonThreeMethods1979}, Stratified Monte Carlo~\citep{haberModifiedMonteCarloQuadrature1966}, or the low-discrepancy sequences used in Quasi-Monte Carlo methods~\citep[chapter 9]{asmussenStochasticSimulationAlgorithms2007}, might yield more accurate surrogate models.
However, a uniform grid structure has many advantages.
First, it is often more convenient to manually inspect a dataset with such a structure.
Second, it allows for the use of certain efficient interpolation methods such as splines, for which a uniform grid structure is mandatory.
Also, the regularity of the grid can be exploited to accelerate nearest-neighbor and linear interpolations.
In this work, we aim to perform a fair comparison between interpolation and ANN regression methods and, thus, we restrict the structure of the dataset used for fitting to uniform grids.

Figure~\ref{fig:interpolation} shows a comparison of the aforementioned interpolation methods on the log Rosenbrock function:
\begin{align}
  \label{eq:rosenbrock}
  \log R : \boldx \in \R^2 \mapsto \log \left[
    1 + (1 - x_1)^2 + 100 \, \left(x_2 - x_1^2 \right)^2
  \right],
\end{align}
which is positive and admits a minimum at $(1,1)$ such that $\log R (1,1) = 0$.
The interpolation methods are fitted on a $7 \times 7$ coarse regular grid on the square $[-1, 1] \times [-0.5, 1.5]$.
They are then evaluated on a $101 \times 101$ much finer grid on the same square.
The accuracy of each method is evaluated using the absolute error (AE) that quantifies how distant the prediction $\fell(\boldx)$ is to the corresponding true value $y_{\ell}$
\begin{align}\label{eq:def_ae}
  \text{AE}\left(\ffull \, ; (\boldx, y_{\ell}) \right) =
  \vert \fell(\boldx) - y_{\ell} \vert
  .
\end{align}
The absolute error is chosen in this example because it is more intuitive to interpret than other error functions such as the squared error (see Sect.~\ref{sec:nn:learning}) or the Cauchy error (see Sect.~\ref{subsec:outlier_detection}).
In general, the results of such a comparison depend on the choice of the error function.

The fitted models and the associated errors shown in Fig.~\ref{fig:interpolation}.
The set of absolute errors is summarized with its mean and maximum values.
The figure reveals general properties of considered interpolation methods.
The nearest-neighbor interpolation provides a piece-wise constant surrogate model with high errors.
The piece-wise linear and cubic spline yield better accuracies.
RBF interpolation performs best on this synthetic case with respect to both mean and max absolute error, but can be outperformed on other examples, mostly by spline interpolation.
As the grid is coarse, all four methods struggle to reproduce the banana shape of the Rosenbrock function.
In ISM models, such strong and fast variations can correspond to a change of physical regime and are thus of critical importance.

\subsection{Performing regression with neural networks}

By relaxing the constraint of passing exactly through the points of the dataset of precomputed models, the derivation of a surrogate model becomes a regression problem.
In machine learning, a regression problem consists in estimating the function $\widehat{\mathbf{f}}: \R^D \rightarrow \R^L$ that best maps input vectors $\boldx$ to output vectors $\boldy$.
This function $\widehat{\mathbf{f}}$ is learned from a dataset of precomputed models \mbox{$\mathcal{D} = \{ (\boldx_n, \boldy_n) \in \R^D \times \R^L, n=1, \ldots, N \}$}.
In this work, the input vector $\boldx$ corresponds to a vector of physical parameters (e.g., temperature, thermal pressure, volume density) and the output vector $\boldy$ to observables computed by a numerical code (e.g., intensities of specific lines).
To perform this estimation, functions $\ffull$ are parametrized with vectors $\paramNN$.
This parametrization restricts the search to a class of functions.
In the following, functions are sometimes denoted $\ffull_{\paramNN}$ to emphasize this association.
For instance, in linear regression, an affine function $\boldx \mapsto \boldW \boldx + \boldb$ is uniquely described by $\paramNN = (\boldW, \boldb)$.
Given the complexity of ISM numerical models, this class is too restrictive to produce accurate surrogate models, and richer classes are required.

Multiple classes of functions and the associated regression algorithms enable the emulation of complex nonlinear functions from data of precomputed models, such as polynomial functions, $k$-nearest-neighbor regression (used e.g., in~\cite{smirnov-pinchukovMachineLearningacceleratedChemistry2022}), Gaussian process regression~\citep{rasmussenGaussianProcessesMachine2006}, decision trees, and the associated ensemble methods such as random forests (used e.g., in~\cite{bronTracersIonizationFraction2021}) or XGBoost~\citep{chenXGBoostScalableTree2016}, ANNs, and others.
All methods based on decision trees or nearest neighbors yield piece-wise functions, which prevents  a desirable regularity property to be enforced in the surrogate model (e.g.,  continuity or differentiability).
Besides, all the listed algorithms, except ANNs and nearest-neighbor interpolation, can only handle multiple outputs independently, which slows down predictions when the number of outputs is high.
An ANN predicts all outputs at once using a sequence of intermediate computations, which is considerably faster.
In addition, ANNs are known to yield very accurate surrogate models both in theory and in practice.
Finally, an ANN comes with the ability to automatically and efficiently compute the derivative of its outputs with respect to its inputs, using automatic differentiation~\citep{paszke2017automatic}.
Overall, to address the complexity of ISM numerical models, exploit prior knowledge on the regularity of the function to approximate, and efficiently predict all outputs simultaneously, we adopted the rich and versatile class of ANNs.
Below, we introduce this class of functions and then describe the approach used to fit an ANN to a dataset.

\subsubsection{Generalities on neural networks}
\label{sec:nn:generalities}

\begin{figure}[t]
  \centering
  \includegraphics[width=0.98\linewidth]{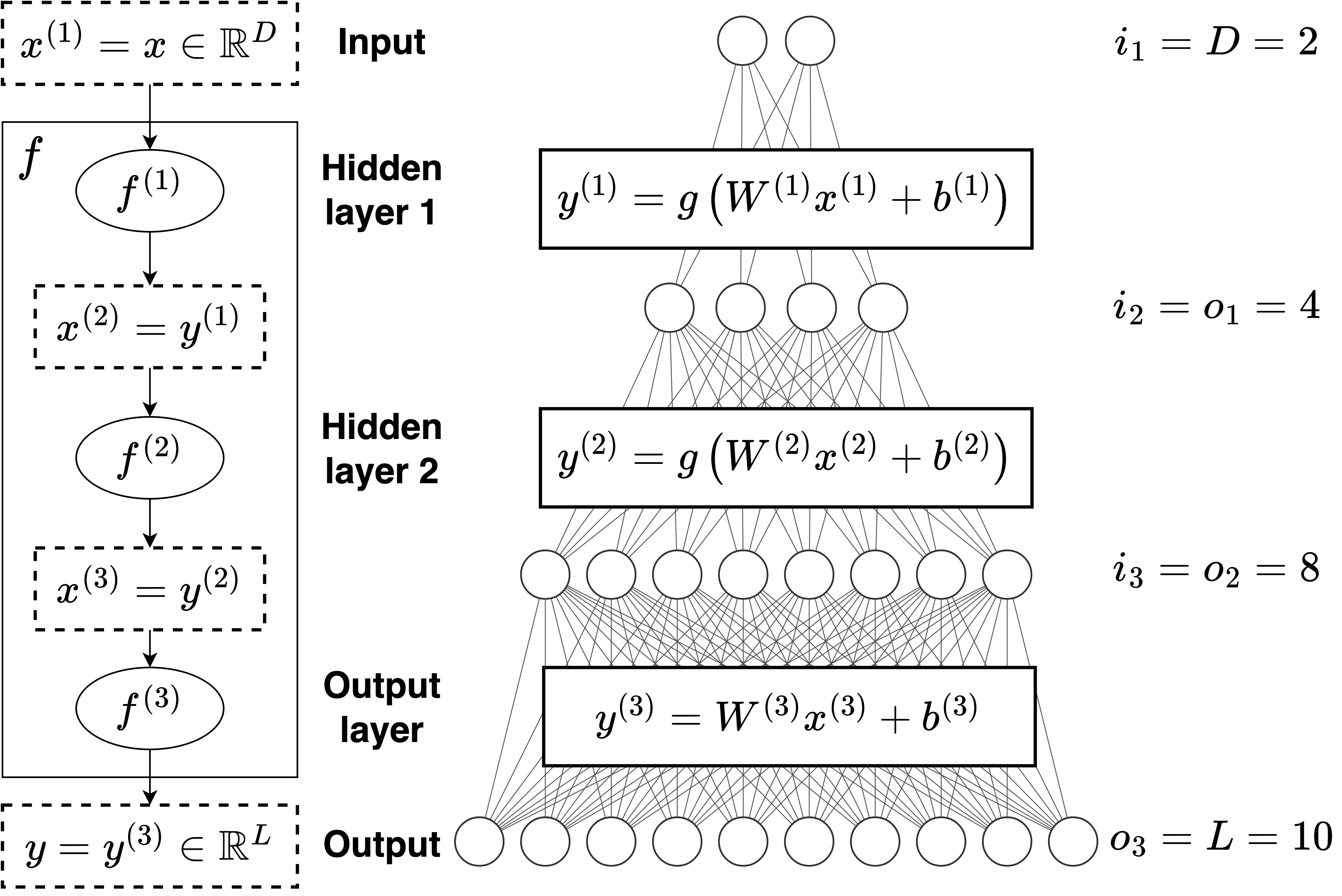}
  \caption{
    Structure of a simple feedforward ANN with $H=2$ hidden layers and a linear layer graph, shown on the left.
  }
  \label{fig:chain-nn}
\end{figure}

The class of mathematical models known as ANNs are  inspired by biological neural systems.
The first ANN was proposed in~\cite{mcculloch_logical_1943} to perform logical operations.
Since then, multiple hardware (e.g., GPU computing) and algorithmic developments (e.g., backpropagation as in~\citealt{rumelhart_learning_1986}) endowed them with the capacity to learn more complex patterns and relationships among the data.
They enjoy fundamental theoretical results.
For different sets of assumptions on the architecture, universal approximation theorems establish that ANNs can approximate almost any continuous function with an arbitrary high level of precision~\citep{hornik_multilayer_1989, leshno_multilayer_1993}.
This class gained widespread popularity after the 2012 ImageNet Challenge, an image classification competition in which an ANN significantly outperformed rival methods~\citep{krizhevsky2017imagenet}.
Nowadays, they are considered a state-of-the-art method for a variety of tasks in vector, image, sound, or text processing across multiple scientific or industrial fields, including astrophysics.
For instance, ANNs have been successfully applied in
exoplanet detection~\citep{shallue_identifying_2018},
Galaxy morphology classification~\citep{huertas-company_catalog_2015}, ISM magnetohydrodynamic turbulence classification~\citep{peek_androids_2019}, and to approximate ISM numerical models~\citep{grassi_robo_2011,demijollaIncorporatingAstrochemistryMolecular2019,holdshipChemulatorFastAccurate2021,grassi_reducing_2022}.
For a more general review of applications of Machine Learning in astronomy, see~\cite{fluke_surveying_2020}.

Throughout this work, an ANN is considered as a function $\ffull : \boldx \in \R^D \mapsto \boldy \in \R^L$, where $D$ and $L$ are input and output dimensions, respectively.
For a numerical model, $D$ is the number of considered physical parameters, such as thermal pressure or visual extinction, and $L$ is the number of predicted observables, for instance, line intensities.
An ANN is made of $H+1$ intermediate functions $\ffull^{(j)}$, called ``layers''.
Intermediate layers $1 \leq j \leq h$ are called the ``hidden layers'' and the final layer is the ``output layer''.
The $j^{\text{th}}$ layer takes an intermediate vector $\boldx^{(j)} \in \R^{i_j}$ as input and computes an intermediate output $\boldy^{(j)} \in \R^{o_j}$.
The intermediate dimensions $i_j$ and $o_j$ can be chosen arbitrarily, except for $i_1 = D$ and $o_{H+1} = L$.
In a feedforward ANN, connections between layers form an acyclic graph.
The output of a layer $j$ feeds one or more of the next layers $j' > j$, hence the notion of direction in a feedforward ANN.

Figure~\ref{fig:chain-nn} shows the structure of a simple ANN that contains $H = 2$ hidden layers and one output layer.
This ANN takes in input $D=2$ physical parameters and predicts $L=10$ observables.
It is indeed a feedforward ANN as its layer graph is linear, as shown on the left.
The output of one of its layers $j$ is thus the input of the next layer $j+1$, that is to say \mbox{$\boldx^{(j+1)} = \boldy^{(j)}$} and \mbox{$i_{j+1} = o_j$}.
Alternative feedforward architectures with nonlinear layer graph exist, such as residual \mbox{networks~\citep{he_deep_2016}} and dense networks~\citep{huang_densely_2017}.
These architectures include skip connections between layers that bypass the activation function to preserve original input information and intermediate computations.
However, linear layer graphs are widespread and remain the simplest multilayer architectures for vector classification or regression tasks.
In the rest of this paper, all the ANNs exhibit such architectures, unless otherwise noted.

A hidden layer combines an affine transformation and a nonlinear scalar function~$g^{(j)}$ applied element-wise as follows:
\begin{align}
  \ffull^{(j)} : \boldx^{(j)} \mapsto \boldy^{(j)} = g^{(j)} ( \boldW^{(j)} \boldx^{(j)} + \boldb^{(j)} ),
\end{align}
with
$\boldW^{(j)} \in \R^{o_j \times i_j}$
and
$\boldb^{(j)} \in \R^{o_j}$
the weight matrix and bias vector of the affine transformation, respectively.
The nonlinear scalar function $g^{(j)}$ is called an activation function.
Common activation functions include the sigmoid, hyperbolic tangent, rectified linear units (ReLU), and multiple variants~\citep{nwankpa_activation_2021}.
Choosing different activation functions $g^{(j)}$ for the $H$ hidden layers might lead to better performance but would require training many ANNs.
A unique $g$ is therefore generally set for all hidden layers.

The output layer transforms the outputs of one or more hidden layers into the desired prediction using an affine transformation and an output activation function.
This output activation function depends on the considered problem.
The sigmoid and the softmax functions are usually employed to return probabilities in binary and multiclass classification, respectively.
In regression tasks, the identity function is generally used.

Overall, in a regression context, the architecure of an ANN is uniquely defined by
its layer graph,
an activation function $g$,
a number of hidden layers $H \geq 0$,
and a sequence of sizes representing its layers $(i_j, o_j)_{j=1}^{H+1}$.
The corresponding class of ANNs is parametrized with a vector $\paramNN = ( \boldW^{(j)}, \boldb^{(j)} )_{j=1}^{H+1}$ that can be very high-dimensional, depending on the number of hidden layers $H$ and their sizes $(i_j, o_j)_{j=1}^{H+1}$.
We note that if $g$ is differentiable, so is the full ANN $\ffull$.
The gradient $\nabla_\boldx \ffull$ can then be efficiently evaluated with automatic differentiation techniques~\citep{paszke2017automatic}.

\subsubsection{Fitting a neural network to a dataset}
\label{sec:nn:learning}

In regression, once the class of function is set (here with an ANN architecture), the parameter $\paramNN$ is adjusted so that $\ffull_{\paramNN}$ fits the dataset $\mathcal{D}$ of precomputed models.
A loss function $\mathcal{L}( \ffull \, ; \mathcal{D})$ quantifies the distance between predictions $\fell(\boldx_n)$ and the corresponding true values $y_{n \ell}$.
It is based on an error function, such as the absolute error (AE, Eq.~\ref{eq:def_ae}) or the squared error (SE), as follows:
\begin{align}\label{eq:def_se}
  \text{SE}\left(\ffull \, ; (\boldx, y_{\ell}) \right) =
  \left( \fell(\boldx) - y_{\ell} \right)^2
  .
\end{align}
The loss function summarizes the set of $N \times L$ errors obtained on the dataset $\mathcal{D}$.
The mean is often used for computational efficiency of evaluation and differentiation, yielding, for example, the mean squared error~(MSE) or the mean absolute error~(MAE).
Obtaining the best function $\widehat{\ffull}$ boils down to minimizing the loss function with respect to the parameter $\paramNN$
\begin{align}\label{eq:optim_pb}
  \widehat{\ffull} \in \arg\min_{\paramNN} \mathcal{L} \left(\ffull_{\paramNN} \, ; \mathcal{D} \right)
  .
\end{align}

Problems of the form of Eq.~\ref{eq:optim_pb} rarely admit a closed-form solution.
Furthermore, with ANNs, the loss function $\mathcal{L}(\ffull_{\paramNN} \, ; \mathcal{D})$ is generally not convex and contains multiple saddle points and local minima~\citep[chapter 20]{shalev-shwartz_understanding_2014}.
Such problems can be solved approximately using a meta-heuristic (e.g., genetic algorithms, particle swarm, simulated annealing) when $\paramNN$ is low-dimensional.
As ANNs typically contain at least hundreds of parameters to tune, these methods are prohibitively slow.
In contrast, gradient descent methods are computationally very efficient.
They rely on automatic differentiation to efficiently evaluate the gradient of the loss function $\nabla_{\paramNN} \mathcal{L}$ and on backpropagation~\citep{rumelhart_learning_1986} to efficiently update $\paramNN$.
The stochastic gradient descent algorithm \citep[see, e.g.,][chapter 14]{shalev-shwartz_understanding_2014} accelerates the search by using ``batches'' instead of the full dataset in gradient evaluations.
Preconditioned variants such as \mbox{RMSProp}~\citep{tieleman_lecture_2012} or Adam~\citep{kingma_adam_2017} exploit the local geometry of the loss function to escape from saddle points and further accelerate convergence to a good local minimum.
This optimization procedure is often called ``training phase'' or ``learning phase'' with ANNs, because the network progressively learns from data as the loss function decreases.


\section{Approximating the Meudon PDR code}
\label{sec:meudonpdr}

The Meudon PDR code, selected as a representative ISM model, is presented below.
The datasets used in the comparison between interpolation algorithms and ANNs as well as their preprocessing are described.
Finally, the considered comparison metrics are defined.

\subsection{The Meudon PDR code: A representative ISM model}

The Meudon PDR code\footnote{\url{https://ism.obspm.fr}}~\citep{le_petit_model_2006} is a 1D stationary code that simulates interstellar gas illuminated with a stellar radiation field.
It can simulate the physics and chemistry of a wide variety of environments, such as diffuse clouds, PDRs, nearby galaxies, damped Lyman alpha systems, circumstellar disks, and so on.
It permits the investigation of effects such as the radiative feedback of a newborn star on its parent molecular cloud.

The user specifies physical conditions such as the thermal pressure in the cloud~\Pth, the intensity of the incoming UV radiation field~\Gnaught~(scaling factor applied to the~\citealt{mathis_interstellar_1983} standard field), and the depth of the slab of gas expressed in visual extinctions,~\AV.
The code then iteratively solves large systems of multiphysics equations.
First, the code solves the radiative transfer equation at each position on an adaptive spatial grid, considering absorption in the continuum by dust and in the lines of key atoms and molecules such as H and \HH~\citep{goicoechea2007penetration}.
Then, from the specific intensity of the radiation field, it computes the gas and grain temperatures by solving the thermal balance.
The heating rate takes into account the photoelectric effect on grains as well as cosmic ray heating.
The cooling rate is estimated from the nonlocal thermodynamic equilibrium (non LTE) excitation in the energy levels of the main species by considering radiative and collisional processes as well as chemical formation and destruction.
Additional processes can either heat or cool the gas, such as \HH~heating or gas-grain collisions.
Finally, the chemistry is solved, providing the densities of about 200 species at each position.
About $3\,000$ reactions are considered, both in the gas phase and on the grains.
The chemical reaction network was built combining different sources including data from the KIDA database\footnote{\url{https://kida.astrochem-tools.org/}} \citep{wakelamKIneticDatabaseAstrochemistry2012} and the UMIST database\footnote{\url{http://udfa.ajmarkwick.net/}} \citep{mcelroyUMISTDatabaseAstrochemistry2013} as well as data from articles.
For key photoreactions, we used cross sections from~\cite{heaysPhotodissociationPhotoionisationAtoms2017} and also taken from Ewine van Dishoeck's photodissociation and photoionization database\footnote{\url{https://home.strw.leidenuniv.nl/~ewine/photo/index.html}}.
The successive resolution of these three coupled aspects (radiative transfer, thermal balance, chemistry) is iterated until a global stationary state is reached.
A full run is computationally intensive and typically lasts a few hours.

The code provides density profiles of the chemical species and the temperature profiles of both the grains and the gas.
It also outputs the line intensities emerging from the cloud that can be compared to observations.
As of version 7 (yet to be released), a total of $5\,409$ line intensities are predicted from species such as H$_2$, HD, C$^+$, C, CO, $^{13}$CO, C$^{18}$O, $^{13}$C$^{18}$O, SO, HCO$^+$, OH$^+$,  HCN, HNC, CH$^+$, CN or CS.

We choose to work on the Meudon PDR code because we consider it a representative element of the most complex ISM models.
Multiple complex ISM codes compute numerous observables from a few physical parameters~\citep{ferland_2017_2017,sutherland_mappings_2018}.
The complex physical and chemical processes taken into account in such codes make the relations between the line intensities and the input parameters highly nonlinear and thus challenging to emulate.
Often, ISM numerical models focusing on a subset of physical processes included in the Meudon PDR code, such as radiative transfer and excitation codes, yield simpler relations between observables and input parameters and  might thus be simpler to emulate.

\begin{table}[!t]
  \setlength{\tabcolsep}{1.5pt}
  \begin{center}
    \caption{Input parameters in the Meudon PDR code.}
    \label{tab:input_params}
    \begin{tabular}{lccc} 
      \multicolumn{4}{c}{Free parameters} \\
      \hline
      Parameter & Value & Unit & Grid \\
      \hline
      Gaz thermal pressure, \Pth & $[ 10^5, \, 10^9 ]$ & K cm$^{-3}$ & on log. scale \\
      UV intensity, \Gnaught & $[ 1, \, 10^5 ]$ & (1) & on log. scale \\
      Visual extinction, \AV & $[1, 40]$ & mag & on log. scale \\
      Inclination angle, \paramAngle & $[0, 60]$ & deg & on lin. scale \\
      ~\\
      \multicolumn{4}{c}{Fixed parameters} \\
      \hline
      Parameter & Value & Unit & Note \\
      \hline
      Cosmic ray ionization rate & $10^{-16}$ & s$^{-1}$ per H$_2$ & (2), (3)\\
      Dust extinction curve & Galaxy & \ldots & (4) \\
      $R_V$ & 3.1 & \ldots & (4) \\
      $N_H / E(B-V)$ & $5.8 \times 10^{21}$ & cm$^{-2}$ & (5) \\
      Mass grain/mass gas & 0.01 & \ldots & \ldots \\
      Grain size distribution & $\propto a^{-3.5}$ & \ldots & (6) \\
      Min grain radius & $10^{-7}$ & cm & \ldots \\
      Max grain radius & $3 \times 10^{-5}$ & cm & \ldots \\
    \end{tabular}
    \vspace{1mm}
    \tablebib{
      (1)~\Gnaught~is the scaling parameter relative to the interstellar standard radiation field from~\cite{mathis_interstellar_1983};
      (2)~\cite{le_petit_h3_2004};
      (3)~\cite{indriolo_h3_2007};
      (4)~\cite{fitzpatrick_analysis_2007};
      (5)~\cite{bohlin_survey_1978};
      (6) The distribution of grain radius $a$ is a power law~\citep{mathis_size_1977}.
    }
  \end{center}
\end{table}

\subsection{Dataset generation}
\label{subsec:dataset}

In this work, we restrict ourselves to constant pressure models as they appear to better reconstruct observations for typical PDRs~\citep{marconi_near_1998,lemaire_high_1999,allers_h2_2005,goicoechea_compression_2016,joblin_structure_2018,wu_constraining_2018}.
We approximated the code with respect to the $D = 4$ input parameters that are most relevant for inference~\citep{wu_constraining_2018,palud_estimating_nodate}.
The three main ones are the thermal pressure,~\Pth, the scaling factor,~\Gnaught,~of the interstellar standard radiation field and the size of the slab of gas measured in total visual extinction,~\AV.
As in~\cite{wu_constraining_2018}, we consider a wide variety of environments with $\Pth \in [10^5, \, 10^9]$ K cm$^{-3}$, $\Gnaught \in [1, \, 10^5]$ and $\AV \in [1, \, 40]$ mag.
The Meudon PDR code computes line intensities for multiple angles~\paramAngle~between the cloud surface and the line of sight.
In the Meudon PDR code, this angle~\paramAngle~can cover a $[0, 60]$ deg interval.
A face-on geometry corresponds to $\paramAngle = 0$ deg and $\paramAngle = 60$ deg is the closest to an edge-on geometry.
To enable analyses of PDRs with known edge-on geometry such as the Orion Bar~\citep{joblin_structure_2018}, this angle is added to the considered physical parameters.
Table~\ref{tab:input_params} details the values of the main input parameters and of other parameters, fixed at standard values from the literature.

We generated two datasets of Meudon PDR code evaluations to assess the approximation quality of the Meudon PDR code: a training set and a test set\footnote{Both datasets can be found in \url{https://ism.obspm.fr/files/ArticleData/2023_Palud_Einig/2023_Palud_Einig_data.zip}}.
The training set is used to fit all surrogate models.
It contains $N_\text{train} = 19\,208$ points, structured as a $14 \times 14 \times 14 \times 7$ uniform grid on $(\log_{10} \Pth, \, \log_{10} \Gnaught, \, \log_{10} \AV, \, \paramAngle)$.
This uniform grid structure is chosen to simplify the outlier identification procedure (see Sect.~\ref{subsec:outlier_detection}) and to include spline interpolation in the comparison.
We note that all the other considered methods, and ANNs in particular, do not require such a structure for the training dataset and using other dataset structures generated, for instance, with Latin hypercube sampling~\citep{mckayComparisonThreeMethods1979}, stratified Monte Carlo~\citep{haberModifiedMonteCarloQuadrature1966}, or low discrepancy sequences used in Quasi-Monte Carlo methods~\citep[chapter 9]{asmussenStochasticSimulationAlgorithms2007} might improve accuracy.
The Meudon PDR code predicts line intensities that are strictly positive and span multiple decades.
To avoid giving more weight in the regression to lines with high intensities and disregarding the faintest ones, in the following, \mbox{$\boldy \in \R^L$} denotes the log-intensities.
Similarly,~\Pth,~\Gnaught~and~\AV~are considered in log scale.
Even in log scale, the parameters of interest cover intervals with quite different sizes.
For instance, $\log_{10} \Gnaught \in [0, 5]$, while $\log_{10} \AV \in [0, 1.602]$. In other words, \AV~covers an interval more than three times smaller than \Gnaught.
Both interpolation methods and ANN based regression typically suffer from this difference.
The $D$ parameters are thus standardized to have a zero mean and a unit standard deviation.
This simple transformation generally improves accuracy for both interpolation methods and ANNs~\citep[chapter 25]{shalev-shwartz_understanding_2014}.

The test dataset was used to assess the accuracy of surrogate models on data not used in the training step.
It contains $N_\text{test} = 3\,192$ points.
These points were generated with $456$ independent random draws from a uniform distribution on the $(\log_{10} \Pth, \, \log_{10} \Gnaught, \, \log_{10} \AV)$ cube and with a uniform grid of $7$ values on $\paramAngle$.
To ensure consistent preprocessing between the two sets, both the input values $\boldx$ and output values $\boldy$ of the test set undergo the same transformations as for the training set.
In particular, the standardization applied to its input values $\boldx$ relies on the means and standard deviations obtained on the training set, and its output values $\boldy$ are considered in decimal log scale.

Numerical codes may yield numerical instabilities.
In its domain of validity, the Meudon PDR code produces few of them.
However, the considered complex nonlinear physics can also lead to physical bistabilities or multistabilities.
For example, the H$_2$ heating process can produce bistable solutions~\citep{burton_line_1990, rollig_kosma-_2022}.
In such a case, profiles computed by the code, for example, of a species density or of the gas temperature, can oscillate between the possible solutions at each position in the modelled cloud.
The line integrated intensities computed from these profiles can contain errors of up to a factor of $100$ and thus are highly unreliable.
The code being deterministic, an input vector $\boldx$ consistently leads to a unique output vector $\boldy$.
However, in the regions of the parameter space with such multistabilities, variations of intensities can be very chaotic and challenging for a surrogate model to reproduce.
Such chaotic values thus lead to the deterioration of the accuracy of any surrogate model, interpolation, or ANN, thus they should not be used.
Unfortunately, as of today there exists no simple or complete procedure to check the physical validity of a precomputed model of the Meudon PDR code.
With a first scan of the datasets, we remove a few lines that are particularly affected.
The total number of considered lines is therefore reduced from $5\,409$ to $L = 5\,375$.
This simple filter leaves other outliers in the training and test datasets.
Although we observe that these outliers are rare (i.e., less than 1\% expected), we do not have any specific a priori knowledge on their location nor on their exact proportion.
To manually check the validity of each value is unrealistic given the sizes of the two datasets.
The most informative hypothesis we can make on outliers is that if one line in a precomputed model is identified as an outlier, then it is likely for this precomputed model to contain other outliers, especially in the lines of the same species or of isotopologues.
This hypothesis is exploited in the more thorough outlier detection method using an ANN, which is presented and described in Sect.~\ref{subsec:outlier_detection}.

Overall, the Meudon PDR code version to emulate is a function $\ffull: \boldx \in \R^D \mapsto \boldy \in \R^L$, with $D = 4$ and $L = 5\,375$.
We assume the predictions of the Meudon PDR code $\ffull$ to vary continuously with respect to the inputs, except in the case of outliers that should be disregarded.
We also assume $\ffull$ to be differentiable.
In Sect.~\ref{sec:using_nn}, we build our emulators such that they satisfy these regularity properties.

\subsection{Comparison metrics}
\label{subsec:metrics}

Interpolation methods and ANNs are compared on evaluation speed, memory requirements, and approximation accuracy.
We describe here the metrics used for the comparison, regardless of how the surrogate models are defined or trained.

The evaluation speed is measured on the full set of $L$ lines for 1\,000 random points.
The measurements are performed on a personal laptop equipped with a 11th Gen Intel(R) Core(TM) i7-1185G7, with eight logical cores running at \SI{3.00}{\giga\hertz}.
The ANNs and interpolation methods are run on CPU to obtain a meaningful comparison.
Running ANNs on a GPU could further reduce their evaluation times.
The implementations of interpolation methods are from the SciPy Python package, popular in ISM studies~\citep{wu_constraining_2018}.
Nearest-neighbor, linear, and RBF interpolation implementations allow for the evaluation of a vector function at once.
Conversely, the spline interpolation implementation requires looping on the $L$ lines, which is a slow process.
To avoid an unfair comparison, the spline interpolation speeds are not evaluated.

The memory requirements are quantified with the number of parameters necessary to fully describe the surrogate model.
Interpolation methods, for instance, require storing the full training set.
It corresponds to $N_{\text{train}} (D + L) \simeq 1.03 \times 10^8$ parameters.
In Python, these parameters are stored using 64-bit floating-point numbers.
Storing the full grid requires about \SI{1.65}{\giga\byte}.

The accuracies of surrogate models are evaluated on the test set, which contains points that they did not see during training.
To quantify accuracies, we define a new metric called the error factor~(EF).
As line intensities are considered in log-scale, the absolute error (Eq.~\ref{eq:def_ae}) corresponds to the ratio (in log scale) of predicted and true line intensities.
The error factor is this ratio transformed back in linear scale.
For a surrogate model $\ffull$ on a given tuple $(\boldx_n, \boldy_n)$ and line $\ell$, it is expressed as:
\begin{align}\label{eq:error_factor}
  \text{EF} \left(\ffull \, ; (\boldx_n, y_{n \ell}) \right)
  & = \; 10^{\left\vert \fell(\boldx_n) - y_{n \ell} \right\vert}
  = \max \left\{
    \frac{10^{\fell(\boldx_n)}}{10^{y_{n \ell}}}, \;
    \frac{10^{y_{n \ell}}}{10^{\fell(\boldx_n)}}
  \right\},
\end{align}
where both $y_{n\ell}$ and $\fell(\boldx_n)$ are line log-intensities.
As the absolute value ensures positivity in log scale, an error factor is always superior or equal to 1.
It can be expressed in percents using a $100 \times (\text{EF} - 1)$ transformation.
For readability, error factors are displayed in percents when EF $< 2$, that is 100\%.
An error factor that is not in percents is indicated by the multiplication sign. For instance, ``$\times$3'' corresponds to EF $=3$.

The error factor is a symmetrized relative error, as the absolute value also ensures symmetry in log scale.
For small errors, namely, EF $\simeq 1$, it is similar to the standard relative error.
However, for larger errors, the error factor is more relevant in our case.
A standard relative error would return 100\% for a factor of two too high and 50\% for a factor of two too low, while in both cases, \mbox{EF $=2$}.
In the worst case, a relative error of 100\% corresponds to a factor of two too high or a prediction of exactly zero, while \mbox{EF $=2$} in the former case and \mbox{EF $=+\infty$} in the latter.
Minimizing a standard relative error would therefore lead to an under-estimation tendency, which is not the case for the proposed error factor.

When applied to the full test set, the error factor yields a distribution of errors.
This distribution is summarized by its mean, its 99th percentile, and its maximum.
The mean provides an estimation of the average error to expect.
The 99th percentile and maximum provide upper bounds on the error.
The maximum is very sensitive to outliers while the 99th percentile is more robust.
To illustrate this sensitivity of upper bounds, consider a fictional dataset of error factors including 0.5\% of outliers at much higher values.
The maximum is affected by the outliers, which induces a pessimistic bias for the corresponding error upper bound estimation.
The 99th percentile is not significantly affected by the outliers, and provides a more relevant estimator of the actual upper bound of the error factor for this fictional dataset.
This example shows that the choice of percentile is a trade-off based on the expected proportion of outliers.
Lower percentiles (e.g., 90 or 95) underestimate the upper bound on the error factor and percentiles higher than 99.5 would, in turn, be sensitive to outliers like the max.
The training and test sets generated with the Meudon PDR case are expected to contain less than 1\% of outliers.
The 99th percentile is therefore expected to be an estimator of the error upper bound that is robust to outliers.

In current IRAM-30m observations, the relative day-to-day calibration accuracy ranges from 3\% to 10\% \citep[see, e.g.,][]{einig_denoising_2023}.
The absolute flux calibration accuracy for ground based observations is more difficult to estimate but cannot be better than the relative calibration accuracy.
For a surrogate model to be relevant for observations analysis and physical parameter inference, we set the constraint that satisfactory surrogate models must have a mean error factor below 10\%.


\section{Designing and training adapted ANN}
\label{sec:using_nn}

The choice of architecture and training approach of ANNs are now discussed.
In the following, ANNs are trained with the MSE loss function.
In addition, we assume the Meudon PDR code to be differentiable.
To derive an ANN satisfying this constraint, we set the activation function $g$ to the exponential linear unit \citep[ELU,][]{nwankpa_activation_2021}.
Unless explicitly mentioned, our ANNs have $H = 3$ hidden layers of equal size.
This choice may not be optimal.
A hyperparameter optimization step could improve the network performance, but would require a validation dataset and the training of many networks.
As the results of Sect.~\ref{sec:results} will show, this step is not necessary to obtain satisfactory results.

The specificities inherent in ISM models such as the Meudon PDR code, namely, the presence of outliers and the unusual dimensions of the problem, very few inputs to predict many outputs.
To address these specificities, the required dedicated strategies are as summarized here and described
in the subsections that follow:
1) we apply an outlier removal procedure;
2) we cluster lines to obtain homogeneous groups simpler to emulate with separate networks;
3) to select an adequate size for hidden layers, we resort to a dimension reduction technique;
4) we apply a polynomial transform to augment the input data and thus ease the learning of nonlinearities; and
5) finally, we replace the standard ANN architecture by a dense architecture exploits values in intermediate layers to re-use intermediate computations.

\subsection{Removing outliers from the training set}
\label{subsec:outlier_detection}

\begin{figure}[!t]
  \centering
  \includegraphics[width=0.495\textwidth]{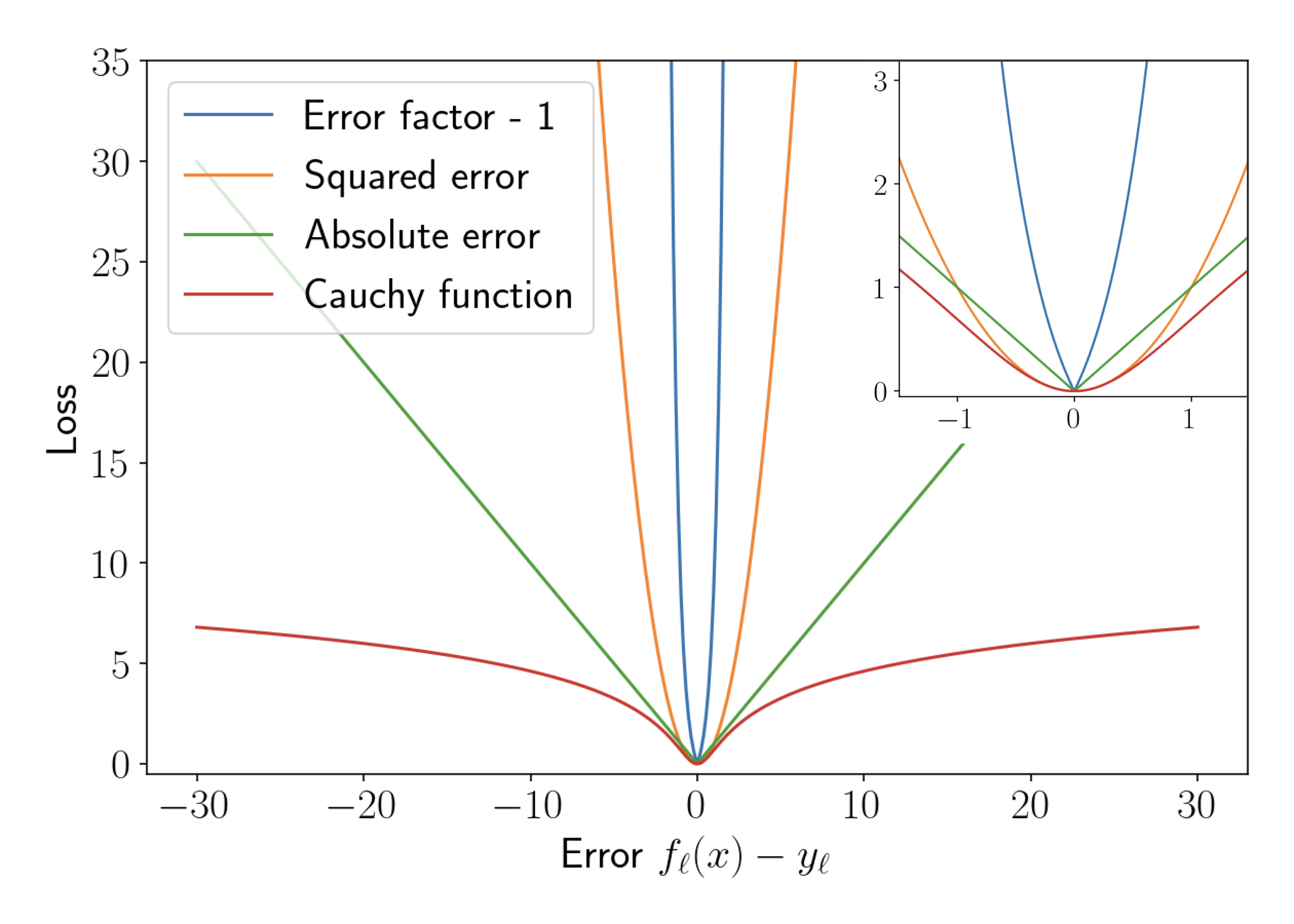}
  \caption{
    Graph of different loss functions.
    As with the line intensities, errors are in decimal log scale.
    An error of $30$ thus corresponds to a factor of $10^{30}$ between predicted and true intensities.
    Since some line intensities range from $10^{-50}$ to $10^{-2}$, this kind of very high error can occur, especially early in the training phase.
  }
  \label{fig:loss_functions}
\end{figure}

Outliers that come from either numerical instabilities or physical bistabilities or multistabilities can be found in both the training and test sets, as described in Sect.\ref{subsec:dataset}.
With a loss function such as the MSE, outliers in the training set greatly deteriorate the quality of a fitted neural network.
Performing regression in presence of outliers is thus a crucial topic in machine learning.
Multiple methods exist for nonlinear regression~\citep{rousseeuw_robust_1987}.
We resort to the method proposed in~\cite{motulsky_detecting_2006}.
This method fits a statistical model to the training set with a strategy robust to outliers.
Then, the training points with largest errors are reviewed.
Identified outliers are removed, and a new statistical model is fitted to the cleaned training set.
To avoid any risk of biasing our analysis towards optimistic results, we do not remove any value from the test set.

For this first fit, we resort to an ANN designed as described at the introduction of Sect.~\ref{sec:using_nn}.
The size of hidden layers is fixed with the dimension reduction strategy that will be described in Sect.~\ref{subsub:pca}.
We also include the polynomial transform of the input, to be described presented in Sect.~\ref{subsub:poly}.
For specific outlier removal step, this fit is performed using the Cauchy loss function (CL):\
\begin{align}
  \label{eq:cauchy_loss}
  \text{CL} \left(\ffull \, ; (\boldx_n, y_{n \ell}) \right)
  & = \; \log \left[
    1 +  \left(\fell(\boldx_n) - y_{n \ell} \right)^2
  \right]
  .
\end{align}
Figure~\ref{fig:loss_functions} shows how the squared error (Eq.~\ref{eq:def_se}), the absolute error (Eq.~\ref{eq:def_ae}), the error factor (Eq.~\ref{eq:error_factor}), and Cauchy function penalize errors.
The Cauchy function gives less weight to outliers than the other error functions, which makes it more robust to outliers. 

The review of training points with high errors is performed with a manual procedure.
An instability in a given model of the grid does not affect all lines, as all lines are not emitted in the same spatial regions of the model.
Therefore, we only remove affected lines instead of the full model.
To accelerate this procedure, we exploit similarities between lines are exploited.
For instance, when one water line intensity is identified as an outlier, it is highly likely that most of the water line intensities of the corresponding precomputed model are outliers.
We emphasize that outliers are associated to instabilities or multistabilities.
Physically consistent intensities that are challenging to reproduce (e.g., due  to fast variations in a change of regimes) are not considered as outliers and maintained in the training set.
In total, $71\,239$ values were identified as outliers, making up 0.069\% of the training set.
We note that this outlier identification step is very informative, as it reveals regions of the parameter space that lead to multistabilities.
However, studying these regions is beyond the scope of this paper.
A binary mask matrix $\mathbf{M} = (m_{n \ell})_{n \ell}$ is defined from this review.
It permits to disregard only the identified outliers instead of removing all $L$ lines of precomputed models with at least one outlier.
In this binary mask, $m_{n \ell} = 1$ indicates that $y_{n \ell}$ is an outlier and should not be taken into account, and $m_{n \ell} = 0$ indicates that $y_{n \ell}$ is not an outlier.
Elements of the training set \mbox{$(\boldx_n, \boldy_n) \in \R^D \times \R^L$} are augmented with corresponding binary mask vectors \mbox{$\boldm_n \in \{0, 1\}^L$}.
On the one hand, ANNs can easily take this mask into account for training by computing the loss function and its gradient on non-masked values only.
In the following, a masked version of the MSE relying on the binary mask $\mathbf{M}$ is used when this outlier removal step is taken into account.

Existing implementations of interpolation methods, on the other hand, lack flexibility to handle such a mask during the fit.
As some points of the grid are removed for some lines, the spline interpolation cannot be applied on the masked training set.
Nearest-neighbor, piece-wise linear and RBF interpolation methods can be applied but would require line by line fits and predictions, as outliers don't occur for the same training points $\boldx$ for all lines.
Such a line-by-line manipulation would be extremely slow with a Python implementation.
To present a somewhat meaningful comparison between ANNs and interpolation methods on the masked dataset, the masked values are imputed.
This imputation relies on a line by line fit of an RBF interpolator with linear kernel.
Masked values are replaced by interpolations computed from available data points.
Interpolation methods are then fitted with this imputed training set.

\subsection{Exploiting correlations between line intensities}
\label{subsec:correlations}

\begin{figure}[!t]
  \centering
  \includegraphics[width=0.999\linewidth]{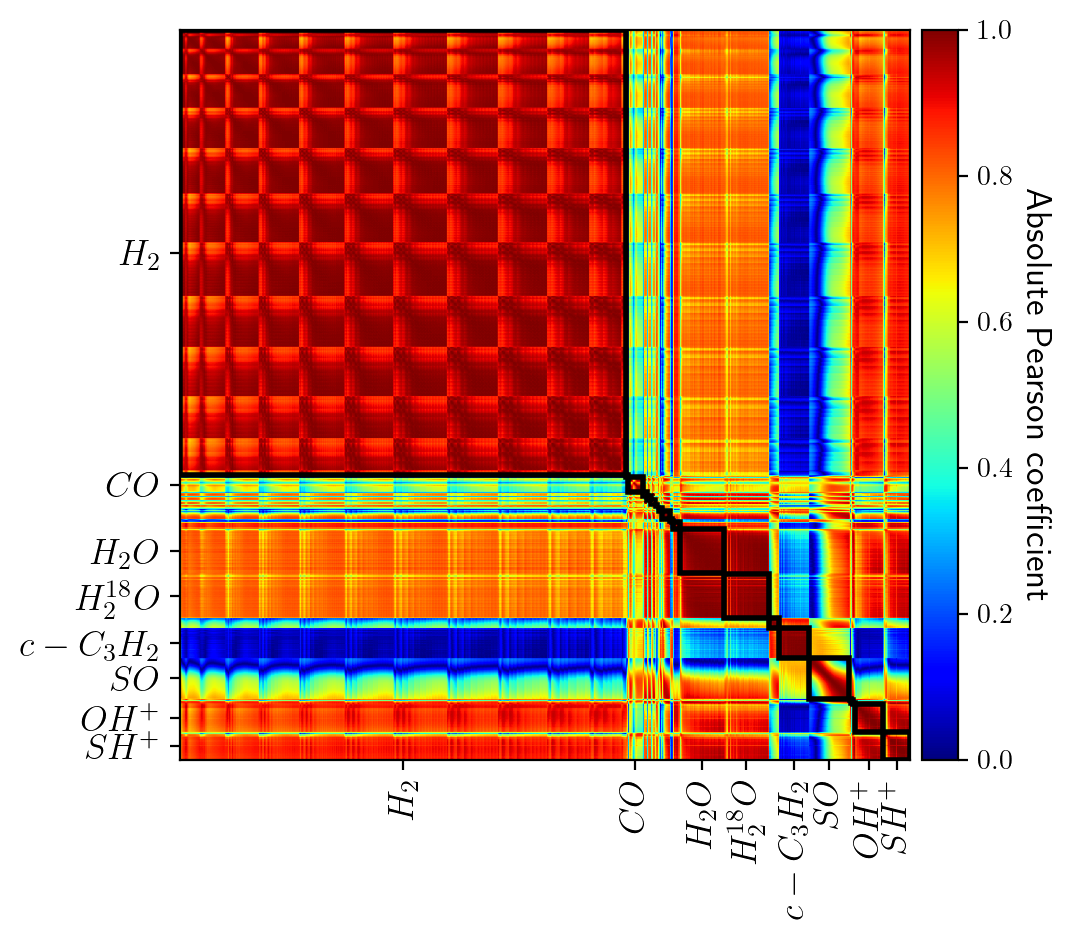}
  \caption{
    Meudon PDR code correlations among the $L = 5\,375$ predicted lines from 27 chemical species, shown with the $L\times L$ matrix of absolute Pearson correlation coefficients.
    A value of exactly 1 for two lines means that there exists an exact affine relationship between their log-intensities.
    The black squares on the diagonal group lines from a common chemical species.
    For readability reasons, only the names of species with more than 100 lines predicted by the Meudon PDR code are displayed.
  }
  \label{fig:pearson_matrix}
\end{figure}

Line intensities computed by the Meudon PDR code come from the radiative de-excitation of energy levels.
While non-local effects are accounted for in the radiative transfer, the excitation of many lines is affected to a large extent by local variables such as the gas temperature or density.
As a result, high correlations between some lines are expected.
Figure~\ref{fig:pearson_matrix} shows the $L \times L$ matrix of absolute Pearson correlation coefficients, with lines grouped by molecule.
We indeed find some strong correlations.
In particular, lines from the same species are often highly correlated, especially for water isotopologues and molecular hydrogen.
However, some species produce lines that are not correlated.
For instance, high energy lines of SO have a very small correlation with low energy lines, as the corresponding submatrix has a diagonal shape.
Finally, some lines from different species are highly correlated, such as OH$^+$, SH$^+$, and H$_2$.
Handling the $L$ lines independently, as in the interpolation methods, ignores these correlations in the line intensities.
We exploit these  correlations with two strategies: a line clustering and a dimension reduction.

\subsubsection{Line clustering to divide and conquer}
\label{subsubsec:outputs_clustering}

\begin{figure}[!t]
  \centering
  \includegraphics[width=0.995\linewidth]{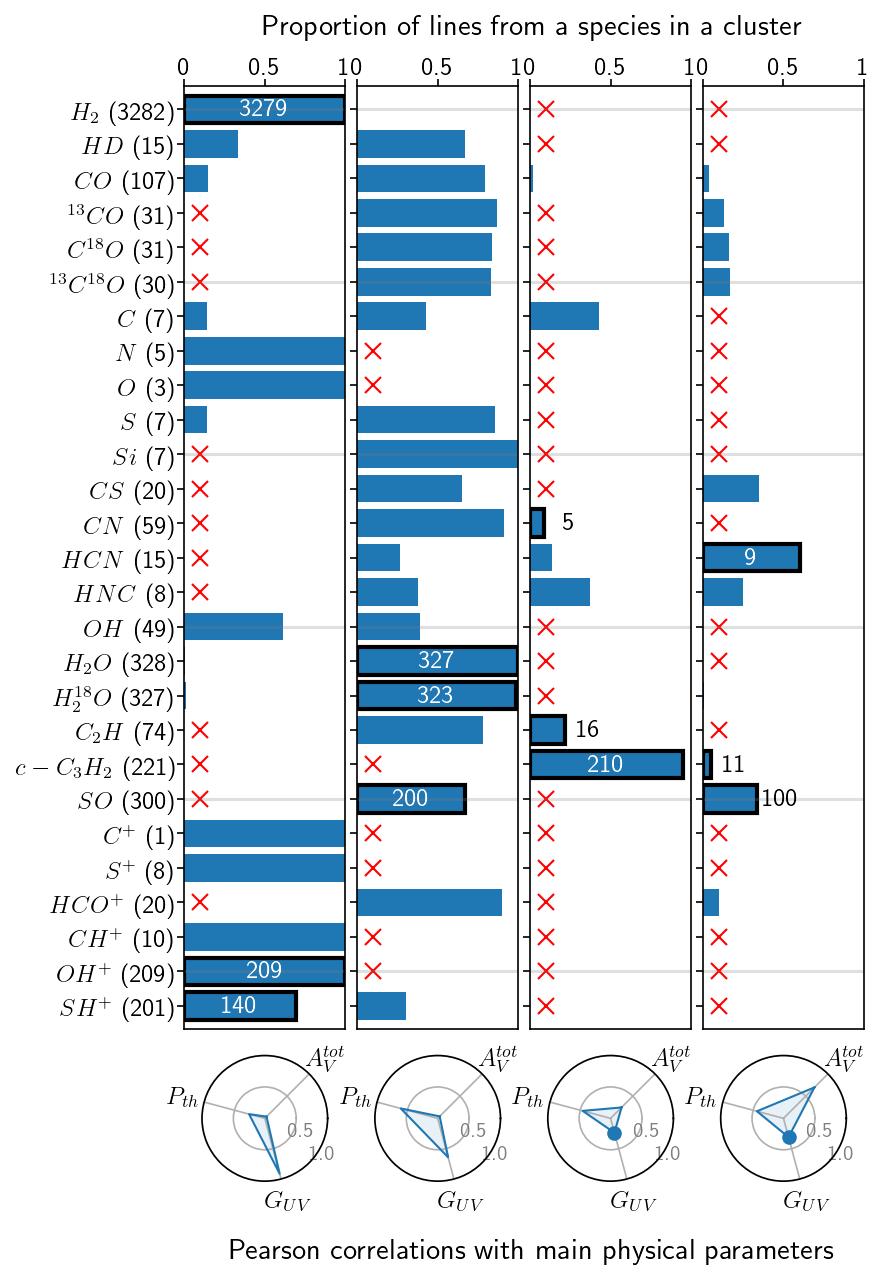}
  \caption{
    Description of the four obtained line clusters.
    Top row: Composition of each cluster.
    Each bar indicates the proportion of lines of a species in a cluster.
    The red crosses correspond to exactly zero line.
    For each cluster, the three species with most lines and the corresponding number of lines are highlighted.
    Bottom row: Pearson correlation of the most representative line of each cluster with the three main physical parameters.
    The most representative line of a cluster is defined as the line with the highest average correlation with the other lines.
    A round marker at a vertex indicates a negative correlation.
  }
  \label{fig:cluster_composition}
\end{figure}

Some clusters of highly correlated lines appear in Fig.~\ref{fig:pearson_matrix}.
These clusters are not simply related to the line carrier.
We derive clusters of lines automatically from the correlation matrix using the spectral clustering algorithm~\citep[chapter 22]{shalev-shwartz_understanding_2014}.
Spectral clustering defines clusters such that lines from the same cluster are as similar as possible and such that lines from different clusters are as different as possible.
It relies on similarity measures (such as a Pearson correlation), while most clustering algorithms are distance-based.
We set the number of clusters to the value that maximizes the ratio of intra to inter-cluster mean correlations.

Figure~\ref{fig:cluster_composition} presents the four clusters we obtained.
The mean intra and inter-cluster correlations are $0.895$ et $0.462$, respectively, while the mean correlation among all lines is $0.73$.
The obtained clusters contain $3\,712$, $1\,272$, $241$, and $150$ lines, respectively.
This imbalance between clusters comes from the imbalance between molecules.
For instance, H$_2$ corresponds to $3\,282$ lines, that is 61\% of the lines computed by the Meudon PDR code, and these lines all are highly correlated as shown in Fig.~\ref{fig:pearson_matrix}.
Appendix~\ref{app:clusters} provides a more complete description of the content of these four clusters.
With this approach, an ANN is trained for each cluster.

\subsubsection{Using PCA to set the size of the last hidden layer}
\label{subsub:pca}

A second and complementary approach to exploit these correlations is to hypothesize that a vector $\boldy$ with the $L$ line intensities can be compressed to a vector of size $\widetilde{L} < L$ with a limited loss of information.
Formally, we hypothesize that the line intensities $\boldy$ live in a subspace of dimension $\widetilde{L} < L$, where $\widetilde{L}$ can be estimated using a dimension reduction algorithm.
We resort to a principal component analysis~(PCA)~\citep[chapter 23]{shalev-shwartz_understanding_2014} on the training set, which performs compressions using only affine transformations.
We obtain that the compression of all $L = 5\,375$ lines, with only $\widetilde{L} \simeq 1\,000$ principal components leads to a decompression with mean error factor below $0.1$\% on the training set, which confirms our hypothesis.

In an ANN such that $D \ll L$, most parameters belong to the last hidden layer, as illustrated in Fig. \ref{fig:chain-nn}.
The size of this layer is thus critical to obtain a good accuracy.
Too large a layer might lead to overfitting, while too small a layer could not capture the nonlinearities of the dataset.
In regression, this last hidden layer applies an affine transformation.
We therefore set its size to the estimated dimension $\widetilde{L}$.
To predict the $\widetilde{L}$ intermediate values of the last hidden layer, which are then used to predict all $L$ line intensities, the first two hidden layers are set with the same size.

For the networks trained on the four clusters of lines obtained in Sect.~\ref{subsubsec:outputs_clustering}, the size of the last hidden layer is also set to the minimum number of principal components that ensures a decompression with mean error factor below 0.1\% on the training set.
The obtained sizes $\widetilde{L}$ are approximately $500$ (\mbox{about $13$\%} of the $L = 3\,712$ lines of the cluster), $350$ \mbox{(about $28$\%)}, $100$ \mbox{(about $41$\%),} and $75$ \mbox{($50$\%)}, respectively.
As the bigger clusters are the most homogeneous, they have the smallest ratio $\widetilde{L} / L$ of subspace dimension $\widetilde{L}$ with the total number of lines $L$.
The number of parameters necessary to describe four small specialized ANNs is thus greatly reduced in comparison to a single larger general network.

\subsection{A polynomial transform for learning nonlinearities}
\label{subsub:poly}

The nonlinearities in the Meudon PDR code make the approximation task challenging.
In an ANN, nonlinearities come from the activation function $g$.
However, learning meaningful and diversified nonlinearities is difficult with few hidden layers.
Conversely, an ANN with numerous layers can lead to overfitting and requires more time for evaluations and memory for storage.
Preprocessing the physical parameters $\boldx$ with a variety of pre-defined nonlinear functions eases this learning task, while maintaining a small network architecture.
We chose to apply a polynomial transform $P_p$ which replaces the input vector $\boldx$ of a dimension $D$ with an input vector containing all monomials computed from the $D$ entries of degree up to $p$.
For instance, for $D = 3$ and $p = 2$, \mbox{$\boldx = (x_1, x_2, x_3)$} is replaced with \mbox{$P_2(\boldx) = (x_1,\,x_2,\,x_3,\,x_1^2,\,x_2^2,\,x_3^2,\,x_1 x_2,\,x_1 x_3,\,x_2 x_3)\in\R^9$}.
For $D = 4$ and $p = 3$, we have \mbox{$P_3(\boldx) \in \R^{34}$}.
This approach is classic in regression~\citep{ostertagova_modelling_2012} but less common in ANNs.

It is well known in polynomial regression that a high maximum degree $p$ can lead to overfitting~\citep[chapter 11]{shalev-shwartz_understanding_2014}.
The analysis of the physical processes indicates that the gas structure and emission properties depend on control quantities combining \Gnaught, $n_H$ (or \Pth) and \AV.
For instance, $\Gnaught / n_H$ is known to play an important role in PDRs~\citep{sternberg_h_2014}.
It is therefore important to consider monomials combining these three physical parameters.
In contrast, the angle~\paramAngle~is assumed to have a simpler role in the model.
To avoid overfitting, we choose the minimum value that combines the three parameters, $p=3$, and thus consider the polynomial transforms $P_3$.
This transformation is applied to the input variables after the preprocessing step described in Sect.~\ref{subsec:dataset} (log scale for \Pth, \Gnaught~and~\AV, and standardization of the $D=4$ parameters).
It is implemented as an additional first fixed hidden layer.
The gradient $\nabla_\boldx \ffull$ can thus be efficiently evaluated with automatic differentiation methods.

\subsection{Dense networks to reuse intermediate computations}
\label{subsubsec:dense}

\begin{figure}[t]
  \centering
  \includegraphics[width=0.98\linewidth]{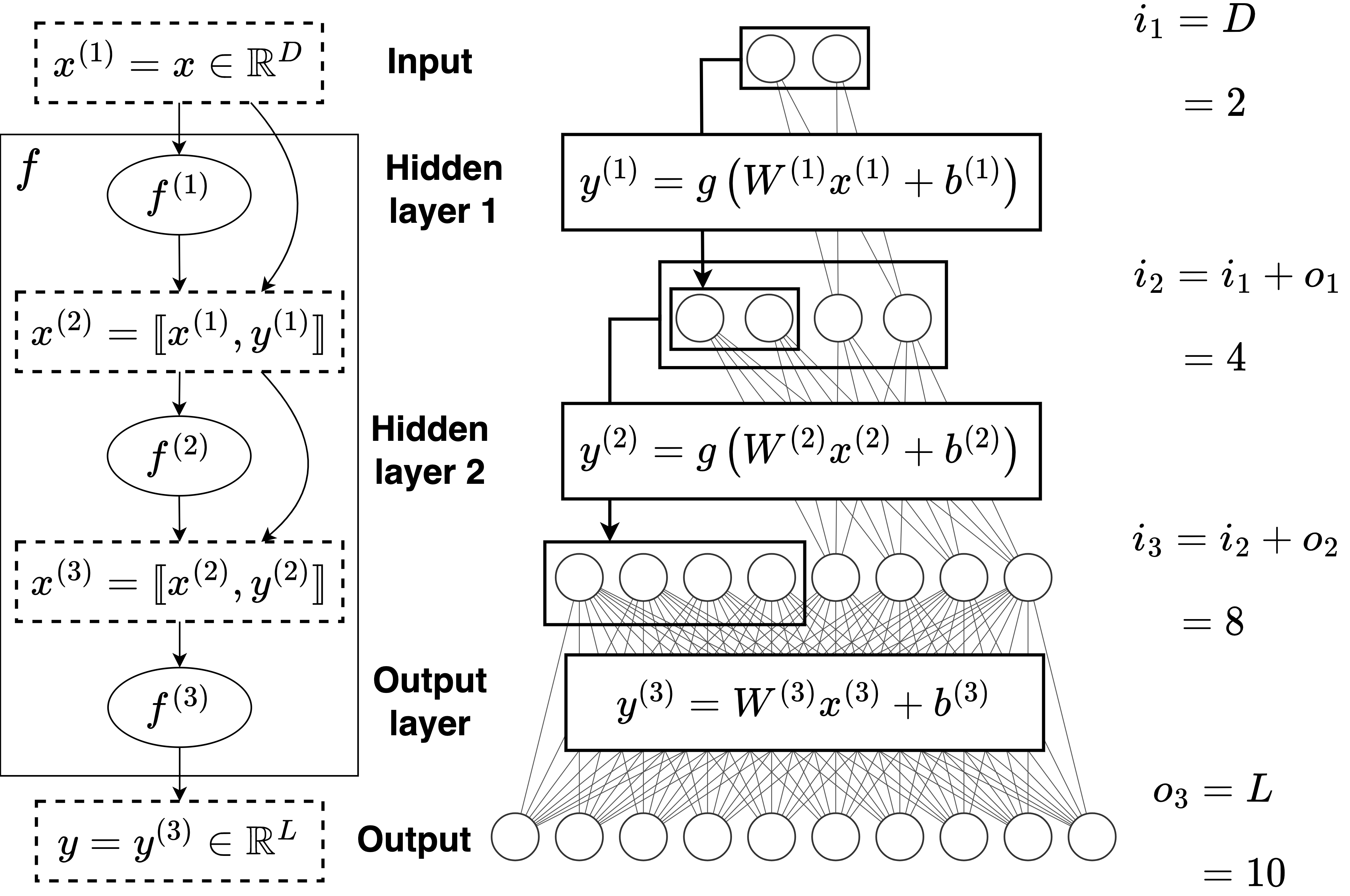}
  \caption{
    Structure of a dense ANN, with \mbox{$H=2$} hidden layers and the same sequence of layer input sizes $(i_j)_{j=1}^{H+1}$ used to illustrate  the feedforward architecture in Fig.~\ref{fig:chain-nn}.
  }
  \label{fig:dense-nn}
\end{figure}

The fully connected ANNs architecture considered so far (shown in Fig.~\ref{fig:chain-nn}) is widely used in the deep learning community.
However, this architecture struggles to maintain input information in hidden layers, as it is transformed in nonlinear activation functions.
It might therefore fail to reproduce very simple relationships.
For instance, the intensity of UV-pumped lines of H$_2$ is highly correlated with~\Gnaught.
Using~\Gnaught~directly to predict intensities of such lines thus might be more relevant than passing it through nonlinear transformations.
This architecture also struggles to pass gradient information all the way back to the first hidden layers.
This phenomenon, known as gradient vanishing, might lead to largely suboptimal trained networks.
The recent residual~\citep{he_deep_2016} and dense architectures~\citep{huang_densely_2017} address these two issues.
We used the dense architecture for our regression problem.

A dense architecture is a special type of feedforward architecture where the input of a layer $j+1$ is the concatenation of the input and output vectors of the previous layer $j$:
$\boldx^{(j+1)} = [\![ \boldx^{(j)}, \boldy^{(j)} ]\!]$.
This architecture focuses on reusing intermediate values in hidden layers and can thus reduce the number of parameters to train.

Figure~\ref{fig:dense-nn} illustrates this dense architecture for a simple ANN with $H=2$ hidden layers and the same sequence of layer input sizes $(i_j)_{j=1}^{H+1}$ used to illustrate the standard feedforward architecture in Fig.~\ref{fig:chain-nn}.
The output sizes $o_j$ of hidden layers are much smaller with the dense architecture, as the input of layer $j$ concatenates the input and output of layer $j-1$.
The weight matrices $\boldW^{(j)}$ of hidden layers are thus much smaller as well, which reduces the total number of parameters to train.
By lowering the number of parameters to learn while providing the same number of inputs to the output layer, this architecture limits overfitting risks.

As the number of parameters per layer is reduced, we define ANNs with $H = 9$ hidden layers, which is six more layers than in the proposed networks with the standard architecture, yet still with a similar number of parameters.
By definition, the size of the hidden layers in a dense architecture is strictly increasing, as the size $i_{j+1}$ of a layer input is the sum $i_j + o_j$ of the input and output sizes of the previous layer.
The network is set so that the input $i_{j+1}$ of a layer $j+1$ is $50$\% larger than the input of the previous layer $i_{j}$.
With this geometric progression and the polynomial transform $P_3$, the input of the output layer contains $1\,296$ neurons, which is $29.6$\% larger than the recommendation from PCA obtained in Sect.~\ref{subsub:pca}.
However, out of these $1\,296$ neurons, $34$ correspond to the input values, $17$ to the output of the first hidden layer, $25$ to the output of the second hidden layer, and so on.
In other words, though the input of the output layer contains more neurons for the considered dense ANN than the PCA recommendation, a majority of these neurons are the result of fewer transformations.

When using this dense architecure strategy with the clustering approach, four dense networks with $H = 9$ hidden layers are designed.
The size of the last hidden layer is also set to a slightly larger value than the PCA recommendation.
The geometric progressions of these four networks are set to 35\%, 30\%, 15\%, and 10\%, respectively.


\section{Results on the Meudon PDR code}
\label{sec:results}

\begin{table}[t]
  \setlength{\tabcolsep}{3pt}
  \centering
  \caption{
    %
    Performance of interpolation methods and of the proposed ANNs, with and without the removal of outlier from the training set.
  }
  \label{tab:results_interpolation}
  \begin{tabular}{ccc|ccccc}
    \multicolumn{3}{c|}{\multirow{2}{*}{Method}}
    & \multicolumn{3}{c}{Error factor}
    & Memory
    & Speed \\
    & &
    & mean
    & 99th per.
    & max
    & (\SI{}{\mega\byte})
    & (\SI{}{\milli\second}) \\
    \hline
    \multirow{10}{*}{\rotatebox[origin=c]{90}{No outlier removal}}
    & \multicolumn{2}{c|}{near. neighbor}
    & $\times$13.1
    & $\times$11.3
    & $\times$3e5
    & $1\,650$
    & 62 \\
    & \multicolumn{2}{c|}{linear}
    & 15.7
    & $\times$2.3
    & $\times$143
    & $1\,650$
    & 1.5e3 \\
    \cline{2-8}
    & \multirow{3}{0.5em}{\rotatebox[origin=c]{90}{spline}}
    & linear
    & 15.7
    & $\times$2.3
    & $\times$144
    & $1\,650$
    & \ldots \\
    & & cubic
    & 11.2
    & $\times$2.2
    & $\times$122
    & $1\,650$
    & \ldots \\
    & & quintic
    & 19.1
    & $\times$2.9
    & $\times$304
    & $1\,650$
    & \ldots \\
    \cline{2-8}
    & \multirow{3}{0.5em}{\rotatebox[origin=c]{90}{RBF}}
    & linear
    & 10.2
    & 96.8
    & $\times$99
    & $1\,650$
    & 1.1e4 \\
    & & cubic
    & 10.4
    & $\times$2.1
    & $\times$112
    & $1\,650$
    & 1.1e4 \\
    & & quintic
    & 10.9
    & $\times$2.1
    & $\times$118
    & $1\,650$
    & 1.1e4 \\
    \cline{2-8}
    & \multirow{2}{0.5em}{\rotatebox[origin=c]{90}{ANN}}
    & R
    & 7.3
    & 64.8
    & \textbf{$\times$81}
    & \textbf{118}
    & \textbf{12} \\
    & & R+P
    & \textbf{6.2}
    & \textbf{49.7}
    & $\times$84
    & \textbf{118}
    & 13 \\[0.5mm]
    \hline
    \hline
    \multirow{13}{*}{\rotatebox[origin=c]{90}{Outlier removal on training set}}
    & \multicolumn{2}{c|}{near. neighbor}
    & $\times$13.1
    & $\times$11.6
    & $\times$3e5
    & $1\,650$
    & 62 \\
    & \multicolumn{2}{c|}{linear}
    & 15.9
    & $\times$2.4
    & $\times$143
    & $1\,650$
    & 1.5e3 \\
    \cline{2-8}
    & \multirow{3}{0.5em}{\rotatebox[origin=c]{90}{spline}}
    & linear
    & 15.9
    & $\times$2.4
    & $\times$144
    & $1\,650$
    & \ldots \\
    & & cubic
    & 11.1
    & $\times$2.2
    & $\times$120
    & $1\,650$
    & \ldots \\
    & & quintic
    & 20.0
    & $\times$2.7
    & $\times$285
    & $1\,650$
    & \ldots \\
    \cline{2-8}
    & \multirow{3}{0.5em}{\rotatebox[origin=c]{90}{RBF}}
    & linear
    & 10.3
    & 97.3
    & $\times$97.5
    & $1\,650$
    & 1.1e4 \\
    & & cubic
    & 10.5
    & $\times$2.0
    & $\times$106
    & $1\,650$
    & 1.1e4\\
    & & quintic
    & 10.9
    & $\times$2.0
    & $\times$114
    & $1\,650$
    & 1.1e4\\
    \cline{2-8}
    & \multirow{5}{0.5em}{\rotatebox[origin=c]{90}{ANN}}
    & R
    & 5.1
    & 42.0
    & \textbf{$\times$32.8}
    & 118
    & 12 \\
    & & R+P
    & 5.5
    & {42.3}
    & {$\times$41}
    & 118
    & 13 \\
    \cline{3-8}
    & & R+P+C
    & {4.9}
    & 44.5
    & $\times$44
    & 51
    & 14 \\
    & & R+P+D
    & \textbf{4.5}
    & \textbf{33.1}
    & $\times$33.8
    & 125
    & \textbf{11} \\
    & & R+P+C+D
    & 4.8
    & 37.9
    & $\times$37.6
    & \textbf{43}
    & 14
    \end{tabular}
    \tablefoot{
      Evaluation speeds are measured on the full set of $L$ lines for 1\,000 random points.
      The measurements are performed on a personal laptop equipped with eight logical cores running at \SI{3.00}{\giga\hertz}.
      Error factors are evaluated on the test set.
      For neural network architectures,
      C stands for a line clustering and specialist networks,
      D for a dense architecture,
      P for a polynomial transform
      and R for the design of the last hidden layer using PCA.
    }
  %
\end{table}

Here, we compare ANNs designed and trained with the proposed strategies with interpolation methods with respect to accuracy, memory, and speed.
Table~\ref{tab:results_interpolation} shows the results of the comparison.
It is divided in two halves.
The first presents models trained on the raw training set, while the second contains models trained on the cleaned training set (using the outlier detection procedure of Sect.~\ref{subsec:outlier_detection}).
In each half, the results of interpolation methods are first listed, followed by ANNs combining one or more of the presented strategies.

\subsection{Performance analysis}

The proposed ANNs outperform all interpolation methods on all aspects by a large margin: they are between $100$ and $1\,000$ times faster than reasonably accurate interpolation methods and between $14$ and $38$ times lighter in terms of memory.
Interpolation methods handle the prediction of $L$ lines as $L$ independent operations, while ANNs handle the $L$ lines at once, which is much faster.
Interpolation methods require storing the full training set that contains 103 million 64-bit floating point numbers, that is to say, \SI{1.65}{\giga\byte} in size.
In contrast, ANNs use shared intermediate values in hidden layers to predict all lines, which limits redundant computations and effectively compresses the dataset.
They can thus be fully described with between 2.7 and 7.8 million parameters, that is to say, between \SI{43}{\mega\byte} and \SI{118}{\mega\byte}.
Finally, the proposed ANNs are roughly twice as accurate as the best interpolation methods on average and between two and three times as accurate with respect to the 99th percentile.
Overall, the proposed ANNs are the only surrogate models that yield a mean error factor lower than 10\% and thus that are suited to a corison with actual observations.

\subsection{Removing outliers is crucial}

When the outlier removal step is not applied, the distribution of errors is highly skewed for all surrogate models.
For interpolation methods, the 99th percentile reveals that around 99\% of the predictions correspond to errors lower than a factor of two.
For the best ANN (R+P), it reveals that 99\% of the errors are lower than 49.7\%.
However, for all methods, the maximum error is at least 80 times higher than the 99th percentile and reaches unacceptable values.
An inspection of the highest errors reveals that they are close to training points with outliers, which indicates that these outliers significantly deteriorate the fit.

After removing outliers from the training set, interpolation methods do not show average accuracy improvement.
Only slight improvements can be observed on the 99th percentile and maximum EF, especially for the RBF interpolation methods.
Replacing outliers with interpolated values is therefore not relevant to derive surrogate models based on interpolation methods in this case.
In contrast, the two ANNs trained both with and without the outlier removal step (R and R+P) show consequent improvements.
With outlier removal, the mean EF decreased from 7.3\% and 6.2\% to 5.1\% and 5.5\%, respectively.
Similarly, the 99th percentile dropped from 64.8\% and 49.7\% to 42\% and 42.3\%.
Finally, the maximum error is reduced by more than a factor of two.
These important improvements demonstrate the interest of filtering outliers from the training set before training ANNs.

\subsection{Combining polynomial transform with dense network or line clustering}

The polynomial transform improves the accuracy in presence of outliers in the training set, but then causes it to deteriorate after the outlier removal step.
It provides flexibility to learn abrupt nonlinearities caused by outliers.
However, with outliers removed, the function to learn is smoother.
The EF on the masked training set is 1.44\% without the polynomial transform (R) and 0.77\% with it (R+P), while the EF on the test set is lower without the polynomial transform.
This improvement on the training set does not lead to an improvement on the test set, suggesting an overfit.
The polynomial transform therefore requires additional strategies to better reproduce data unused during the training phase.

Both the clustering step and dense architecture, used with the polynomial transform, led to better accuracy.
The surrogate model that exploits the line clustering but not the dense architecture (R+P+C) improves the mean accuracy by $0.2$ percentage points, while requiring $57$\% fewer parameters than the first two networks (R and R+P).
A potential cause of the average error factor improvement is the separation of the trainings of each specialized ANN.
Since H$_2$ lines represent 61\% of all $L$ lines, they dominate the loss function and thus are learned in priority.
To separate them from other clusters might have improved performance on those other clusters.

The surrogate model based on a single network with dense architecture (R+P+D) is the most accurate on average and provides the lowest error upper bound for the robust 99th percentile estimator.
Even with more trainable parameters than the first two networks, it does not overfit.
It is also the fastest model as reusing intermediate values reduces the number of computations.

Finally, combining both line clustering and dense architectures (R+P+C+D) yields the lowest memory usage with only 2.7 million parameters, that is \SI{43.2}{\mega\byte}, which is 38 times lighter than for interpolation methods.
It also provides very good accuracy, both on the average and for the upper bounds.

Overall, a dense architecture and the line clustering effectively limit overfitting and thus perform better on data unseen during the training phase.
The line clustering leads to the lightest models regarding memory requirements, while the dense architecture leads to the most accurate models.


\section{Conclusion}
\label{sec:conclusion}

The interpretation of observations of atomic and molecular tracers in galactic and extragalactic ISM requires comparison with state-of-the-art astrophysical models to infer physical conditions.
Such inference procedure requires numerous evaluations of the numerical model, which is particularly the case for Bayesian approaches.
Inference on large observations maps, which are becoming more and more common, further relies on many evaluations.
The ISM models are often too slow to perform such inference and are generally approximated using interpolation methods run on grids of precomputed models.
These interpolation approaches induce errors that are seldom quantified in the literature.
Besides, these methods can have high evaluation time and memory costs.

In this work, the general problem of deriving a fast, accurate and memory-light surrogate model for a time-consuming ISM numerical model has been addressed.
The proposed approach has been assessed in the case of the Meudon PDR code, a state-of-the-art ISM code.
In this work, four common families of interpolation methods (nearest-neighbor, linear, spline, and RBF) are compared to specifically designed ANNs.
We find that ANNs outperform all interpolation methods by a large margin in terms accuracy, speed, and memory usage.

Attaining this performance level for ISM models requires addressing their specificities.
First, ISM models usually predict many observables (e.g., line intensities of many species) from few parameters (e.g., gas density or temperature), which is unusual in ANN applications -- except in the case of ANNs that generate structured data such as images, text, or times series.
Second, due to numerical instabilities or physical bistabilities or multistabilities, such models sometimes produce outliers that harm the training process.
We proposed and combined five strategies to design and train adapted ANNs:
\begin{itemize}
  \item To identify outliers, we first train an ANN with a loss function robust to large errors.
  Training points corresponding to large errors are manually reviewed.
  Identified outliers are removed from the training set.
  \item Lines are clustered into homogeneous subsets that are simpler to emulate: for each cluster one ANN is defined and trained.
  \item A dimension reduction technique (PCA) is used to determine an adequate size of hidden layers.
  \item A polynomial transform of the input physical parameters provides precomputed nonlinearities to the network, which permits the learning of nonlinearities with a limited number of hidden layers.
  \item A dense architecture exploits intermediate computations and thus limits redundant computations.
  Using such an architecture instead of the standard feedforward ANN architecture improves speed and avoids overfitting.
\end{itemize}

With the proposed strategies, ANNs can achieve 4.5\% average accuracy, while the best interpolation method, RBF, attains 10.2\%.
The upper bound on the errors, quantified using their 99th percentile, reach 33.1\% for our ANNs compared to 97\% for the RBF interpolation.
Besides, our ANNs are $1\,000$ times faster than RBF and are more than ten times lighter in terms of memory.
The most accurate model presented in this article (denoted R+P+D in Table~\ref{tab:results_interpolation}) is publicly available\footnote{\url{https://ism.obspm.fr/files/ArticleData/2023_Palud_Einig/2023_Palud_Einig_trained_ANN.zip}}.

Although this paper focuses on an application to the Meudon PDR code, the proposed strategies are general enough to be applicable to many other ISM models.
The fast and accurate ANN emulators obtained in this article enable the performance of fully Bayesian inference on observation maps using the Meudon PDR code, a physically comprehensive model.
Such an approach will be presented in an upcoming paper~\cite{palud_estimating_nodate}.
It will also permit efficient analyses of large observations maps produced by today's instruments (e.g., JWST, ALMA), such as the ORION-B dataset observed by the IRAM 30m~\citep{petyAnatomyOrionGiant2017}.


\begin{acknowledgements}
  This work was partly supported by the CNRS through 80Prime project OrionStat, a MITI interdisciplinary program,
  by the ANR project ``Chaire IA Sherlock'' ANR-20-CHIA-0031-01 held by P. Chainais,
  and by the national support within the {\em programme d'investissements d'avenir} ANR-16-IDEX-0004 ULNE and Région HDF.
  It also received support from the French Agence Nationale de la Recherche
  through the DAOISM grant ANR-21-CE31-0010,
  and from the Programme National ``Physique et
  Chimie du Milieu Interstellaire'' (PCMI) of CNRS/INSU with INC/INP,
  co-funded by CEA and CNES.
  JRG and MGSM thank the Spanish MCINN for funding support under grant PID2019-106110G-100.
  Part of the research was carried out at the Jet Propulsion Laboratory, California Institute of Technology, under a contract with the National Aeronautics and Space Administration (80NM0018D0004). D.C.L. was supported by USRA through a grant for SOFIA Program 09-0015.

\end{acknowledgements}


\bibliographystyle{aa} %
\bibliography{main} %

\begin{appendix}


\section{Content of clusters of lines}
\label{app:clusters}

In this appendix, we describe the content of the four clusters shown in Fig.~\ref{fig:cluster_composition}.
All species have lines distributed in at most three clusters, except for $^{12}$CO that has lines in each of the four clusters.
The CO lines are indexed with two quantum numbers: the rotational number $J$ and the vibrational number $v$.

The first cluster gathers lines that are emitted from the most external UV illuminated layers of the cloud and trace hot chemistry.
It includes all H$_2$ lines but three and, thus, it is the largest.
It also contains all lines from OH$^+$ (209), CH$^+$ ', all eight lines of S$^+$, all five lines of N, all three lines of O and the \SI{158}{\micro\metre} line of C$^+$, and CO transitions with low $J$ values for $v=1-0$ and $v=1-1$. %
It also contains all lines from OH$^+$ (209), CH$^+$ (10), S$^+$ (8), N (5), the fine structure lines of O (3) and C$^+$ (1), and rovibrational lines of CO in the $J$ values for $v=1-0$ and $v=1-1$ ladders.
The line intensities of this cluster are highly and positively correlated to \Gnaught, and not correlated at all with \AV.

The second cluster contains 99\% of the 655 lines from water H$_2 ^{16}$O and its isotopologue H$_2^{18}$O.
It also contains lines from high energy levels for several molecules (HD, CO, $^{13}$CO, C$^{18}$O, $^{13}$C$^{18}$O, HNC, HCN, HCO$^+$, SO, CN, SH$^+$, C$_2$H, OH, and CS), as well as transitions from the neutral atoms C, Si, and S.
Line intensities in this cluster are positively correlated with \Pth~and~\Gnaught~and not at all with \AV.

The third cluster contains mostly c-C$_3$H$_2$ lines, and some C$_2$H lines.
It also includes two transitions of CO with moderate $J$ values at the lowest vibrational level $v=0$ ($J =3-2$ and $J =4-3$).
Its line intensities are overall positively correlated with~\Pth~and~\AV~and negatively correlated with~\Gnaught.

The fourth cluster contains the low energy lines of $^{13}$CO, C$^{18}$O, $^{13}$C$^{18}$O, HNC, HCN, HCO$^+$, SO, and c-C$_3$H$_2$, as well as the lowest temperature transitions of CO ($J =1-0$ and $J =2-1$).
Its line intensities are overall positively correlated with~\Pth, strongly positively correlated with~\AV~and negatively correlated with~\Gnaught.

\end{appendix}

\end{document}